\documentclass[11pt]{article}
\usepackage{epstopdf}
\usepackage{color}
\usepackage{subfigure}
\usepackage{amsmath}
\usepackage{amssymb}
\usepackage{graphicx,color}
\usepackage{cite}
\usepackage{enumerate}
\usepackage{amsthm}
\usepackage{amsfonts,mathrsfs}
\usepackage{geometry}
\usepackage{ifthen}

\begin{document}

\title{Geometry of skew information-based quantum coherence}

\vskip0.1in
\author{\small Zhao-Qi Wu\thanks{Corresponding author. E-mail: wuzhaoqi\_conquer@163.com} $^{1,3}$, Huai-Jing Huang$^{1}$, Shao-Ming Fei\thanks{Corresponding author. E-mail: feishm@cnu.edu.cn} $^{2,3}$,
Xian-Qing Li-Jost$^{3}$\\
{\small\it  1. Department of Mathematics, Nanchang University, Nanchang 330031, P R China} \\
{\small\it  2. School of Mathematical Sciences, Capital Normal University, Beijing 100048, P R China}\\
{\small\it  3. Max-Planck-Institute for Mathematics in the Sciences,
04103 Leipzig, Germany}}

\date{}
\maketitle

\noindent {\bf Abstract} {\small } We study the skew
information-based coherence of quantum states and derive explicit
formulas for Werner states and isotropic states in a set of
autotensor of mutually unbiased bases (AMUBs). We also give surfaces
of skew information-based coherence for Bell-diagonal states and a
special class of $X$ states in both computational basis and in
mutually unbiased bases. Moreover, we depict the surfaces of the
skew information-based coherence for Bell-diagonal states under
various types of local nondissipative quantum channels. The results
show similar as well as different features compared with relative
entropy of coherence and $l_1$ norm of coherence.

\vskip 0.1in


\vskip 0.1in


\vskip0.2in

\noindent {\bf 1. Introduction}

Quantum coherence is an intrinsic character of quantum mechanics which plays significant roles in superconductivity, quantum thermodynamics
and biological processes, while a theoretic framework to quantify the coherence was not formulated until the work of \cite{TB}.
It intrigued great interest in studying quantum coherence from different perspectives and aspects.

A large number of valid coherence measures or coherence monotones
such as relative entropy of coherence, $l_1$ norm of coherence,
robustness of coherence, coherence of formation, max-relative
entropy of coherence, modified trace distance of coherence, skew
information-based coherence, geometric coherence, coherence weight,
affinity distance-based coherence, generalized $\alpha$-$z$-relative
R\'enyi entropy of coherence and various entropic-based coherence
measures have been proposed to quantify quantum coherence
\cite{TB,YUAN1,CN,KB1,CX1,XDY,BC1,CSY,LUO1,LUO2,LUO3,KB2,CX2,CX3,XNZ,YUAN2}.
Average coherence and coherence-generating power of quantum channels
based on different coherence measures have also been extensively
explored \cite{SC,LZ1,LZ2,LZ3,LUO4,PZ1,PZ2,PZ3,PZ4,LZ4}. Moreover,
the interconversion between quantum coherence and quantum
entanglement or quantum correlations are formulated
\cite{AS1,EC1,HJZ,YX,JM,LUO5,KIM,KDW}.

On the other hand, the problem of coherence distillation and
coherence dilution have also been discussed
\cite{AW,EC2,BR,KF,CLL,LL,MJZ,QZ}, together with the no-broadcasting
of quantum coherence \cite{ML,IM}. A complete theory of one-shot
coherence distillation has been formulated in \cite{YUAN3}. Quantum
coherence can also be used to certify quantum memories \cite{TS}.
The quantum coherence among nondegenerate energy subspaces (CANES)
has been shown to be essential for the energy flow in any quantum
system \cite{TM}.

The concept of mutually unbiased bases (MUBs) was raised in quantum
state determinations. It is found to be possible to construct $d+1$
MUBs of the space $\mathbb{C}^d$ if $d$ is a prime power, i.e.,
$d=p^n$, where $p$ is a prime number and $n$ is an integer
\cite{IDI,WKW}. It is still not known yet that what are the maximal
sets of MUBs when the dimension $d$ is a composite number \cite{TD}.
The link between unextendible maximally entangled bases and mutually
unbiased bases has been established \cite{BC2}, and entanglement,
compatibility of measurements and uncertainty relations with respect
to MUBs have been investigated \cite{SPENGLER,DC,CC,SD,AER}.

The geometry of entanglement measures and other correlation measures can provide an intuition towards the quantification of these correlations.
The level surfaces of entanglement and quantum discord for Bell-diagonal states \cite{MDL}, the level surfaces of quantum discord for a
class of two-qubit states \cite{BL}, the geometry of one-way information deficit for a class of two-qubit states \cite{YKW1}, the surfaces
of constant quantum discord and super-quantum discord for Bell-diagonal states \cite{YKW2} have been depicted. Recently, the $l_1$ norm of coherence
of quantum states in mutually unbiased bases has been discussed \cite{YKW3}, and the geometry with respect to relative entropy of coherence and
$l_1$ norm of coherence for Bell-diagonal states has been investigated \cite{YKW4,YKW5}.

In this paper, we calculate skew information-based coherence of
quantum states in mutually unbiased bases for qubit and two-qubit
quantum states, and formulate the corresponding geometries. We
explore the geometry of skew information-based coherence of
two-qubit Bell-diagonal states and $X$ states in both computational
basis and in mutually unbiased bases. We also investigate the
dynamic behavior of the skew information-based coherence under
different quantum channels.

\vskip0.1in

\noindent {\bf 2. Skew information-based coherence in autotensor of
mutually unbiased bases}

\vskip0.1in

Let $\mathcal{H}$ be a $d$-dimensional Hilbert space, and
$\mathcal{B(H)}$, $\mathcal{S(H)}$ and $\mathcal{D(H)}$ be the set
of all bounded linear operators, Hermitian operators and density
operators on $\mathcal{H}$, respectively. Usually, a state and a
channel are mathematically described by a density operator (positive
operator of trace $1$) and a completely positive trace preserving
(CPTP) map, respectively \cite{NC}.

The set of incoherent states, which are diagonal matrices in the
fixed orthonormal base $\{|k\rangle\}^d_{k=1}$ of the
$d$-dimensional Hilbert space $\mathcal{H}$, can be represented as
$$\mathcal{I}=\{\delta\in \mathcal{D(H)}|\delta=\sum_{i}p_i|i\rangle\langle i|,~p_i\geq 0,~\sum_{i}p_i=1\}.$$
Let $\Lambda$ be a CPTP map $\Lambda(\rho)=\sum_{n}K_n\rho
K_n^\dag,$ where $K_n$ are Kraus operators satisfying
$\sum_{n}K_n^\dag K_n=I_{d}$ with $I_d$ the identity operator. $K_n$
are called incoherent Kraus operators if $K_n^\dag \mathcal{I}K_n\in
\mathcal{I}$ for all $n$, and the corresponding $\Lambda$ is called
an incoherent operation.

A well-defined coherence measure $C(\rho)$ shall satisfy the
following conditions \cite{TB}:

$(C1)$ (Faithfulness) $C(\rho)\geq 0$ and $C(\rho)=0$ iff $\rho$ is
incoherent.

$(C2)$ (Convexity) $C(\cdot)$ is convex in $\rho$.

$(C3)$ (Monotonicity) $C(\Lambda(\rho))\leq C(\rho)$ for any
incoherent operation $\Lambda$.

$(C4)$ (Strong monotonicity) $C(\cdot)$ does not increase on average
under selective incoherent operations, i.e.,
$$C(\rho)\geq \sum_{n}p_nC(\varrho_n),$$
where $p_n=\mathrm{Tr}(K_n\rho K_n^\dag)$ are probabilities and
$\varrho_n=\frac{K_n\rho K_n^\dag}{p_n}$ are the post-measurement
states, $K_n$ are incoherent Kraus operators.

For a state $\rho\in \mathcal{D(H)}$ and an observable $K\in
\mathcal{S(H)}$, the {\it Wigner-Yanase} (WY) skew information is
defined by \cite{WY}
\begin{equation}\label{eq1}
I(\rho,K)=-\frac{1}{2}\mathrm{Tr}([\rho^{\frac{1}{2}},K]^2),
\end{equation}
where $[X,Y]:=XY-YX$ is the commutator of $X$ and $Y$.

In an attempt to quantify coherence \cite{GIROLAMI}, Girolami
proposed to use the Wigner-Yanase skew information $I(\rho,K)$ to
quantify coherence, and called it $K$-coherence. Here $K$ is
diagonal in the base $\{|k\rangle\}_{k=1}^d$. More precisely, this
quantity should be considered as a quantifier for coherence of
$\rho$ with respect to the observable $K$ rather than the associated
orthonormal base.

The absence of a reference frame has been proven equivalent to
constrain quantum dynamics by a superselection rule (SSR)
\cite{BARTLETT}, while the ability of a system to act as reference
frame is the quantum resource known as asymmetry or frameness
\cite{BARTLETT}. A $G$-SSR for a quantity $Q$ (supercharge) is
defined as a law of invariance of the state of a system with respect
to a transformation group $G$. In \cite{GIROLAMI}, it is shown that
given a $G$-SSR with supercharge $Q$, the skew information $I(\rho,
Q)=-\frac{1}{2}\mathrm{Tr}([\rho^{\frac{1}{2}},Q]^2)$ satisfies the
criteria identifying an asymmetry measure of the state \cite{GOUR}.
It is worth noting that quantum asymmetry represents the amount of
coherence in the eigenbasis of the supercharge \cite{GOUR}.

The $K$-coherence satisfies $(C1)$ and $(C2)$, but not $(C3)$, as
pointed out in \cite{DUBAI,MARVIAN}. By using the spectral
decomposition of the observable $K$ rather than the observable $K$
itself, the authors in \cite{LUO2} have showed that the
$K$-coherence can be simply modified to be a bona fide measure of
coherence satisfying the above requirements $(C1)$-$(C3)$ (where
they called it partial coherence).

Another way to solve the problem is proposed in \cite{CSY} by
introducing the skew information-based coherence measure defined by
\cite{CSY}
\begin{equation}\label{eq2}
C_I(\rho)=\sum_{k=1}^d I(\rho,|k\rangle\langle k|),
\end{equation}
where $I(\rho,|k\rangle\langle
k|)=-\frac{1}{2}\mathrm{Tr}\{[\rho,|k\rangle \langle k|]\}^2$ is the
skew information of the state $\rho$ with respect to the projections
$\{|k\rangle \langle k|\}_{k=1}^d$. Direct calculation shows that
the coherence measure (\ref{eq2}) can be written as
\begin{equation}\label{eq3}
C_I(\rho)=1-\sum_{k=1}^d\langle k|\sqrt{\rho}|k\rangle ^2.
\end{equation}

In \cite{CSY}, it has been proved that the coherence measure defined
in (\ref{eq2}) satisfies all the criteria $(C1)$-$(C4)$, while the
$K$-coherence does not satisfy $(C4)$ (strong monotonicity). The
coherence measure has an analytic expression and an obvious
operational meaning related to quantum metrology. In terms of this
coherence measure, the distribution of the quantum coherence among
the multipartite systems has been studied and a corresponding
polygamy relation has been proposed. It is also found that the
coherence measure gives the natural upper bounds of quantum
correlations prepared by incoherent operations. Moreover, it is
shown that this coherence measure can be experimentally measured.
Since the skew information-based coherence measure (\ref{eq2}) is of
great significance both theoretically and practically, it is worth
evaluating the measure for classes of quantum states in both
computational basis and mutually unbiased bases, and studying the
geometrical characters.

A set of orthonormal bases
$\{e_k\}=\{|0\rangle_k,|1\rangle_k,\cdots,|d-1\rangle_k\}$ for a
Hilbert space $H=\mathbb{C}^d$ is called MUBs if \cite{IDI,WKW}
$$|_k\langle i|j\rangle_l|=\frac{1}{\sqrt{d}}$$
holds for all $i,j\in \{0,1,\cdots,d-1\}$ and $k\neq l$.
For $d=2$, a set of three mutually unbiased bases is given by
$$e_1=\{e_{11},e_{12}\}=\{|0\rangle,|1\rangle\},$$
$$e_2=\{e_{21},e_{22}\}=\left\{\frac{1}{\sqrt{2}}(|0\rangle+|1\rangle),\frac{1}{\sqrt{2}}(|0\rangle-|1\rangle)\right\},$$
$$e_3=\{e_{31},e_{32}\}=\left\{\frac{1}{\sqrt{2}}(|0\rangle+i|1\rangle),\frac{1}{\sqrt{2}}(|0\rangle-i|1\rangle)\right\}.$$

Let $\{e_k\}=\{|0\rangle_k,|1\rangle_k,\cdots,|d-1\rangle_k\}$ be a
set of mutually unbiased bases. The set $\{a_k\}=\{|i\rangle_k\otimes |j\rangle_k,
i,j=0,1,\cdots,d-1\}$ is called the {\it autotensor of mutually unbiased
bases} (AMUBs) if \cite{YKW3}
$$
|(\langle i|_k\otimes \langle j|_k)(|m\rangle_l\otimes |n\rangle_l)|=\frac{1}{d},~~~i,j,m,n=0,...,d-1,
$$
for $k\neq l$.
The following set is an AMUBs derived from two-dimensional MUBs \cite{YKW3}:
$$\{a_1\}=\{a_{11},a_{12},a_{13},a_{14}\}=\{e_{11}\otimes e_{11},e_{11}\otimes e_{12},e_{12}\otimes e_{11},e_{12}\otimes e_{12}\},$$
$$\{a_2\}=\{a_{21},a_{22},a_{23},a_{24}\}=\{e_{21}\otimes e_{21},e_{21}\otimes e_{22},e_{22}\otimes e_{21},e_{22}\otimes e_{22}\},$$
$$\{a_3\}=\{a_{31},a_{32},a_{33},a_{34}\}=\{e_{31}\otimes e_{31},e_{31}\otimes e_{32},e_{32}\otimes e_{31},e_{32}\otimes e_{32}\}.$$

In general, a two qubit $X$ states can be represented as
\begin{equation}\label{eq3}
\rho^X=\frac{1}{4}(I\otimes I+\mathbf{r}\cdot \sigma\otimes
I+I\otimes \mathbf{s}\cdot \sigma+\sum_{i=1}^3 c_i\sigma_i\otimes
\sigma_i),
\end{equation}
where $\mathbf{r}$ and $\mathbf{s}$ are Bloch vectors. As a special
class of $\rho^X$, for $\mathbf{r}=\mathbf{s}=0$, one obtains the
two-qubit Bell-diagonal states
\begin{equation}\label{eq4}
\rho^{BD}=\frac{1}{4}\left(I\otimes I+\sum_{i=1}^3
c_i\sigma_i\otimes \sigma_i\right),
\end{equation}
where $c_i \in [-1,1]$, $i=1,2,3$.

The density matrix of $\rho^{BD}$ in basis $a_1$ is of the form
$$(\rho^{BD})_{a_1}=\frac{1}{4}\left(\begin{array}{cccc}
         1+c_3&0&0&c_1-c_2\\
         0&1-c_3&c_1+c_2&0\\
         0&c_1+c_2&1-c_3&0\\
         c_1-c_2&0&0&1+c_3\\
         \end{array}
         \right),
$$
and the skew information-based coherence of $(\rho^{BD})_{a_1}$ is
\begin{equation}\label{eq5}
C_I(\rho^{BD})_{a_1}=\frac{1}{4}(2-\sqrt{1-c_1-c_2-c_3}\sqrt{1+c_1+c_2-c_3}-\sqrt{1+c_1-c_2+c_3}\sqrt{1-c_1+c_2+c_3}).
\end{equation}

Similarly, the density matrix of $\rho^{BD}$ in basis $a_2$ and
$a_3$ are given by
$$(\rho^{BD})_{a_2}=\frac{1}{4}\left(\begin{array}{cccc}
         1+c_1&0&0&c_3-c_2\\
         0&1-c_1&c_3+c_2&0\\
         0&c_3+c_2&1-c_1&0\\
         c_3-c_2&0&0&1+c_1\\
         \end{array}
         \right)
$$
and
$$(\rho^{BD})_{a_3}=\frac{1}{4}\left(\begin{array}{cccc}
         1+c_2&0&0&c_3-c_1\\
         0&1-c_2&c_3+c_1&0\\
         0&c_3+c_1&1-c_2&0\\
         c_3-c_1&0&0&1+c_2\\
         \end{array}
         \right),
$$
with the skew information-based coherence
\begin{equation}\label{eq6}
C_I(\rho^{BD})_{a_2}=\frac{1}{4}(2-\sqrt{1+c_1+c_2-c_3}\sqrt{1+c_1-c_2+c_3}-\sqrt{1-c_1-c_2-c_3}\sqrt{1-c_1+c_2+c_3})
\end{equation}
and
\begin{equation}\label{eq7}
C_I(\rho^{BD})_{a_3}=\frac{1}{4}(2-\sqrt{1-c_1-c_2-c_3}\sqrt{1+c_1-c_2+c_3}-\sqrt{1+c_1+c_2-c_3}\sqrt{1-c_1+c_2+c_3}),
\end{equation}
respectively.
We plot the level surfaces of $C_I(\rho^{BD})_{a_1}$  in Figure 1.
\begin{figure}[ht]\centering
\subfigure[] {\begin{minipage}[b]{0.3\linewidth}
\includegraphics[width=1\textwidth]{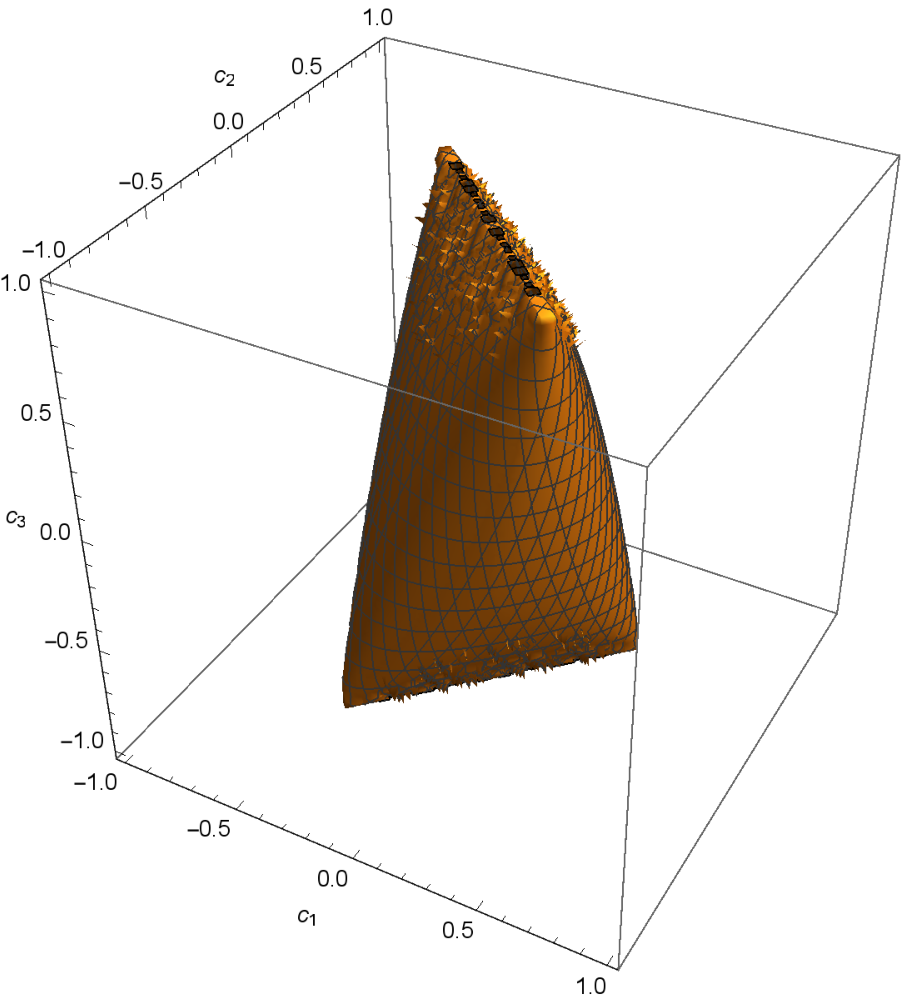}
\end{minipage}}
\subfigure[] {\begin{minipage}[b]{0.3\linewidth}
\includegraphics[width=1\textwidth]{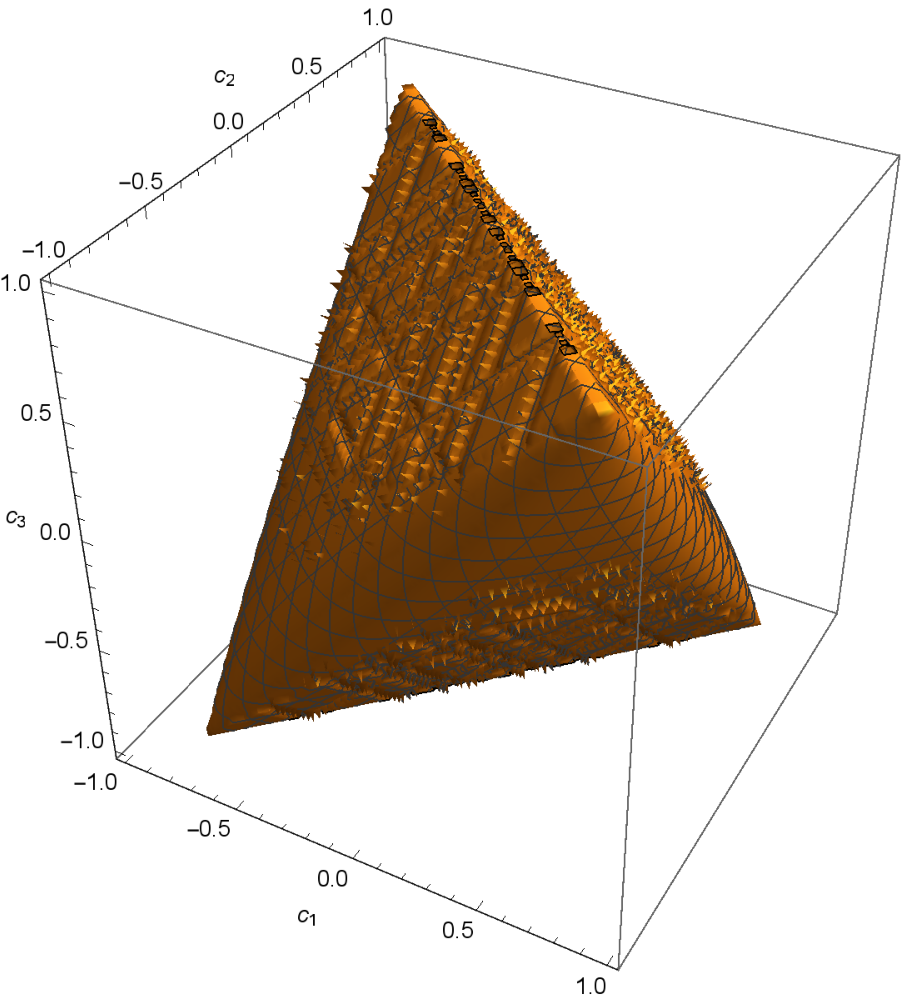}
\end{minipage}}
\subfigure[] {\begin{minipage}[b]{0.3\linewidth}
\includegraphics[width=1\textwidth]{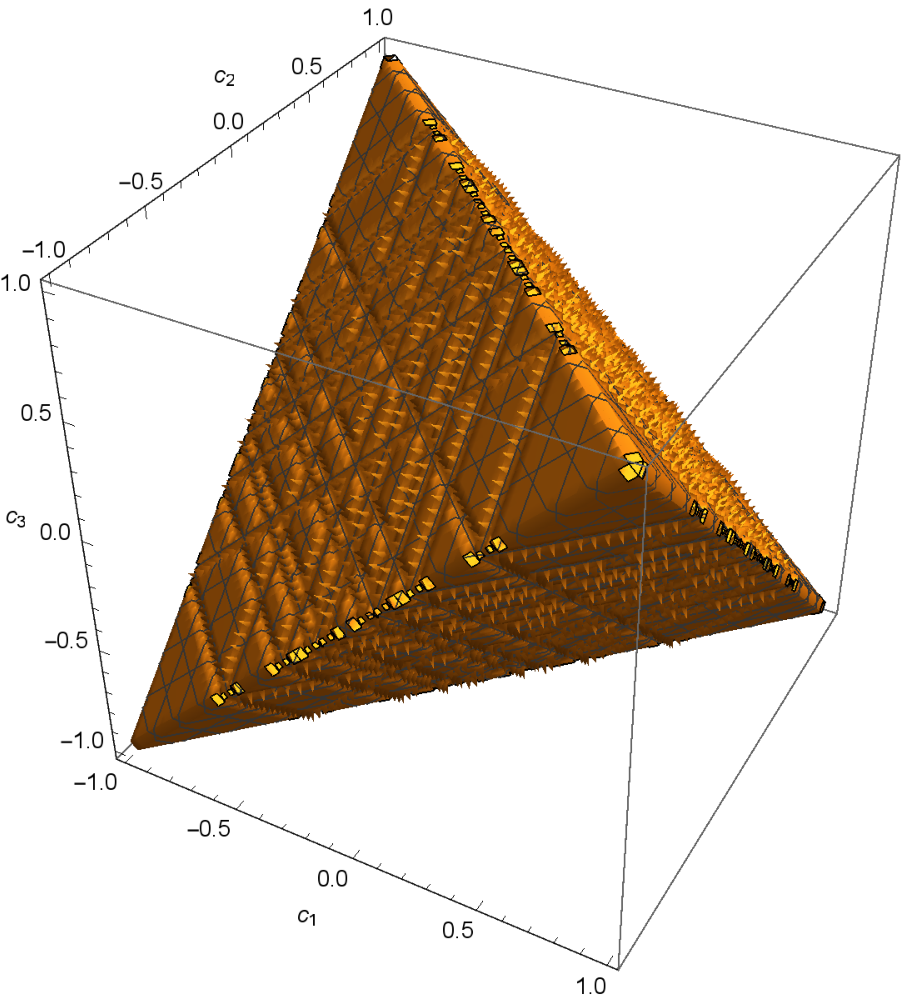}
\end{minipage}}
\caption{Surfaces of constant $C_I(\rho^{BD})_{a_1}$: (a)
$C_I(\rho^{BD})_{a_1}=0.05$; (b) $C_I(\rho^{BD})_{a_1}=0.2$; (c)
$C_I(\rho^{BD})_{a_1}=1$.} \label{fig:MUB1}
\end{figure}

The Bell-diagonal state $\rho^{BD}$ becomes the Werner state
$\rho^W$ if we take $c_1=c_2=c_3=\frac{3}{4}p-1$ ($0\leq p\leq 1$). We have
$$(\rho^W)_{a_i}=\left(\begin{array}{cccc}
         \frac{1}{3}p&0&0&0\\
         0&\frac{1}{6}(3-2p)&\frac{1}{6}(4p-3)&0\\
         0&\frac{1}{6}(4p-3)&\frac{1}{6}(3-2p)&0\\
         0&0&0&\frac{1}{3}p\\
         \end{array}
         \right),~~~i=1,2,3,
$$
and
\begin{equation}\label{eq8}
C_I(\rho^W)_{a_i}=\frac{1}{16}(8-\sqrt{p(48-27p)}-3p),~~~i=1,2,3.
\end{equation}

Taking $c_1=c_3=\frac{4F-1}{3}$, $c_2=-\frac{4F-1}{3}$
($0\leq F\leq 1$), we have the isotropic state $\rho^{iso}$,
$$(\rho^{iso})_{a_i}=\left(\begin{array}{cccc}
         \frac{1}{6}(1+2F)&0&0&\frac{1}{6}(4F-1)\\
         0&\frac{1}{3}(1-F)&0&0\\
         0&0&\frac{1}{3}(1-F)&0\\
         \frac{1}{6}(4F-1)&0&0&\frac{1}{6}(1+2F)\\
         \end{array}
         \right),~~~i=1,2,
$$
and
$$(\rho^{iso})_{a_3}=\left(\begin{array}{cccc}
         \frac{1}{3}(1-F)&0&0&0\\
         0&\frac{1}{6}(1+2F)&0&0\\
         0&0&\frac{1}{6}(1+2F)&0\\
         0&0&0&\frac{1}{3}(1-F)\\
         \end{array}
         \right),
$$
from which we obtain
\begin{equation}\label{eq9}
C_I(\rho^{iso})_{a_i}=\frac{1}{6}(1+2F-2\sqrt{3F(1-F)}),~~~i=1,2,3.
\end{equation}

The $C$-axis stands for $C_I(\rho^W)_{a_i}$ and
$C_I(\rho^{iso})_{a_i}(i=1,2,3)$ in Figure 2(a) and (b),
respectively.
\begin{figure}[ht]\centering
\subfigure[] {\begin{minipage}[b]{0.47\linewidth}
\includegraphics[width=1\textwidth]{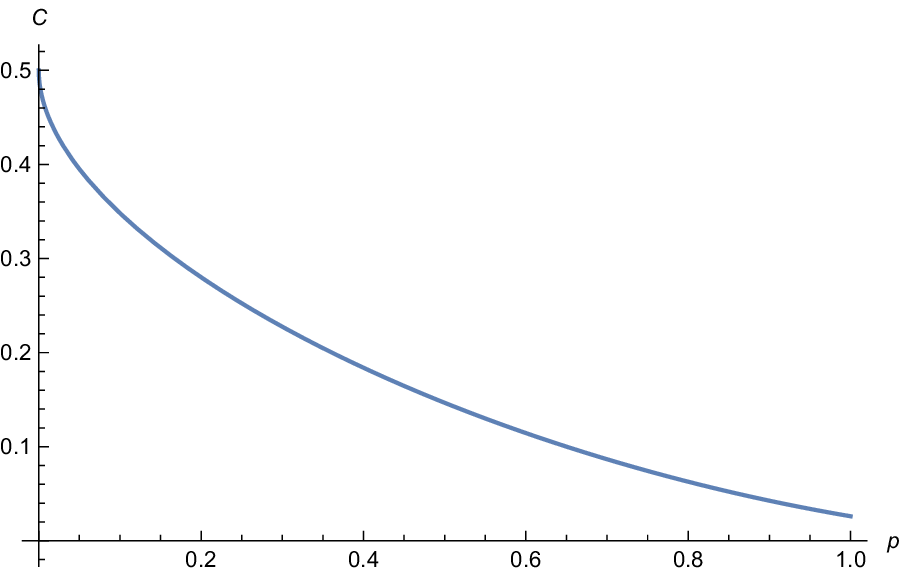}
\end{minipage}}
\subfigure[] {\begin{minipage}[b]{0.47\linewidth}
\includegraphics[width=1\textwidth]{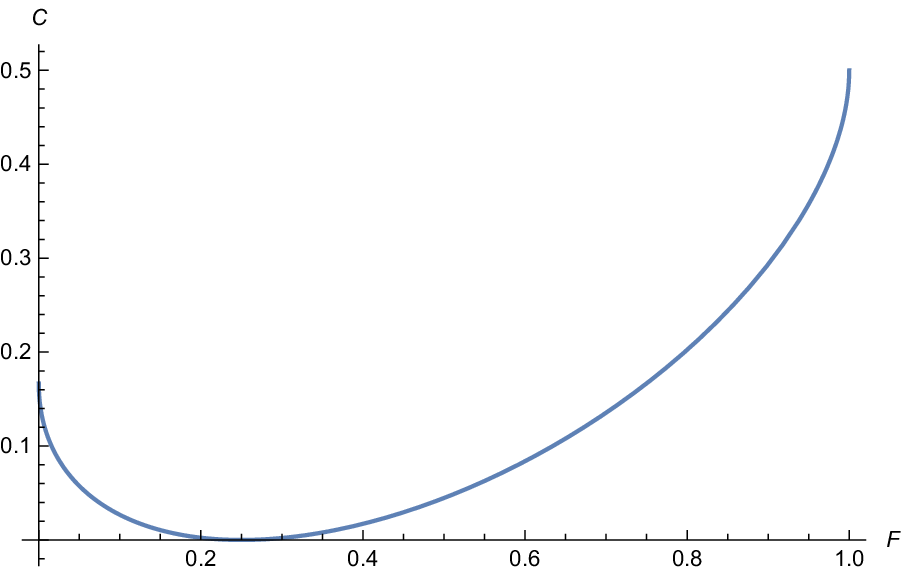}
\end{minipage}}
\caption{(a) $C_I(\rho^W)_{a_i}(i=1,2,3)$ as a function of $p$; (b)
$C_I(\rho^{iso})_{a_i}(i=1,2,3)$ as a function of $F$.}
\label{fig:MUB2}
\end{figure}

Denote the sum of the skew information-based coherence of
Bell-diagonal states in bases $\{a_i\}_{i=1}^3$ by
\begin{equation}\label{eq10}
C_I(\rho^{BD})_{a}=C_I(\rho^{BD})_{a_1}+C_I(\rho^{BD})_{a_2}+C_I(\rho^{BD})_{a_3}.
\end{equation}
In Figure 3, we plot the surfaces of constant $C_I(\rho^{BD})_{a}$ of
Bell-diagonal states $\rho^{BD}$. Comparing Figure 1 with Figure 3, it can be
seen that the volume of the surface expands when both the value of
$C_I(\rho^{BD})_{a_1}$ and $C_I(\rho^{BD})_{a}$ increases.
Moreover, when $C_I(\rho^{BD})_{a_1}$ or $C_I(\rho^{BD})_{a}$ equals
to $1$, both of the surfaces approaches to a tetrahedron.
\begin{figure}[ht]\centering
\subfigure[] {\begin{minipage}[b]{0.3\linewidth}
\includegraphics[width=1\textwidth]{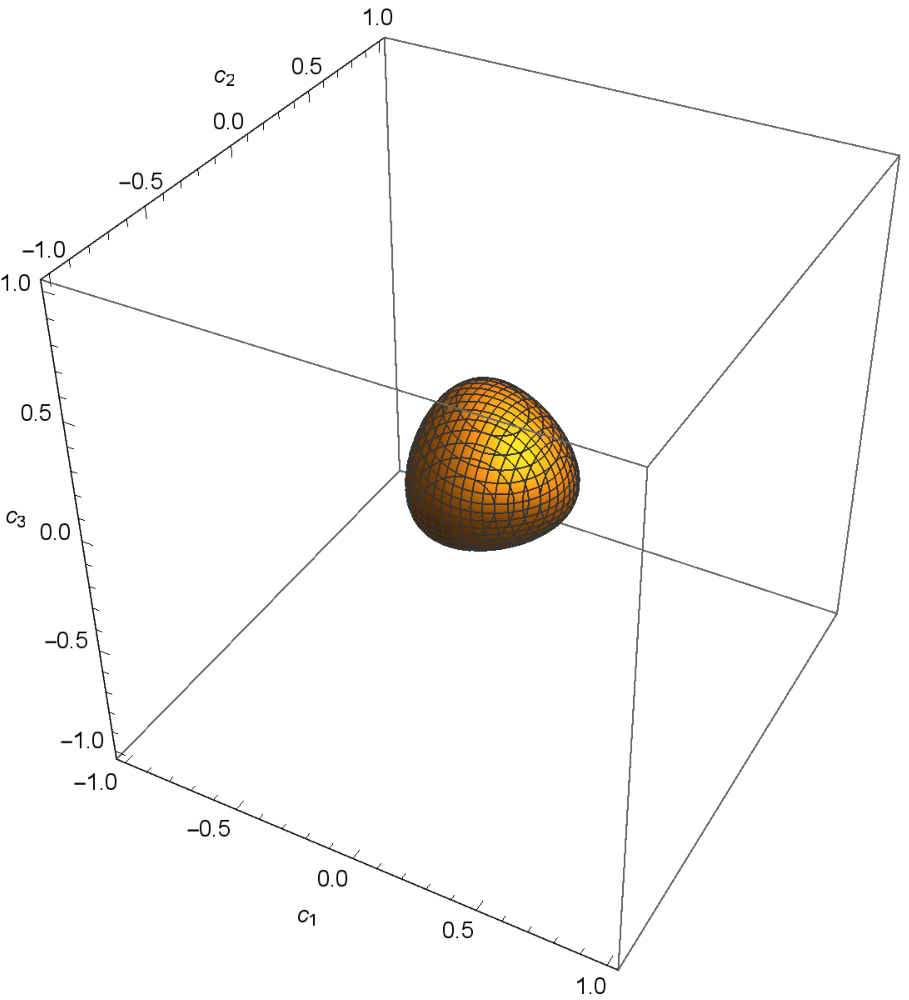}
\end{minipage}}
\subfigure[] {\begin{minipage}[b]{0.3\linewidth}
\includegraphics[width=1\textwidth]{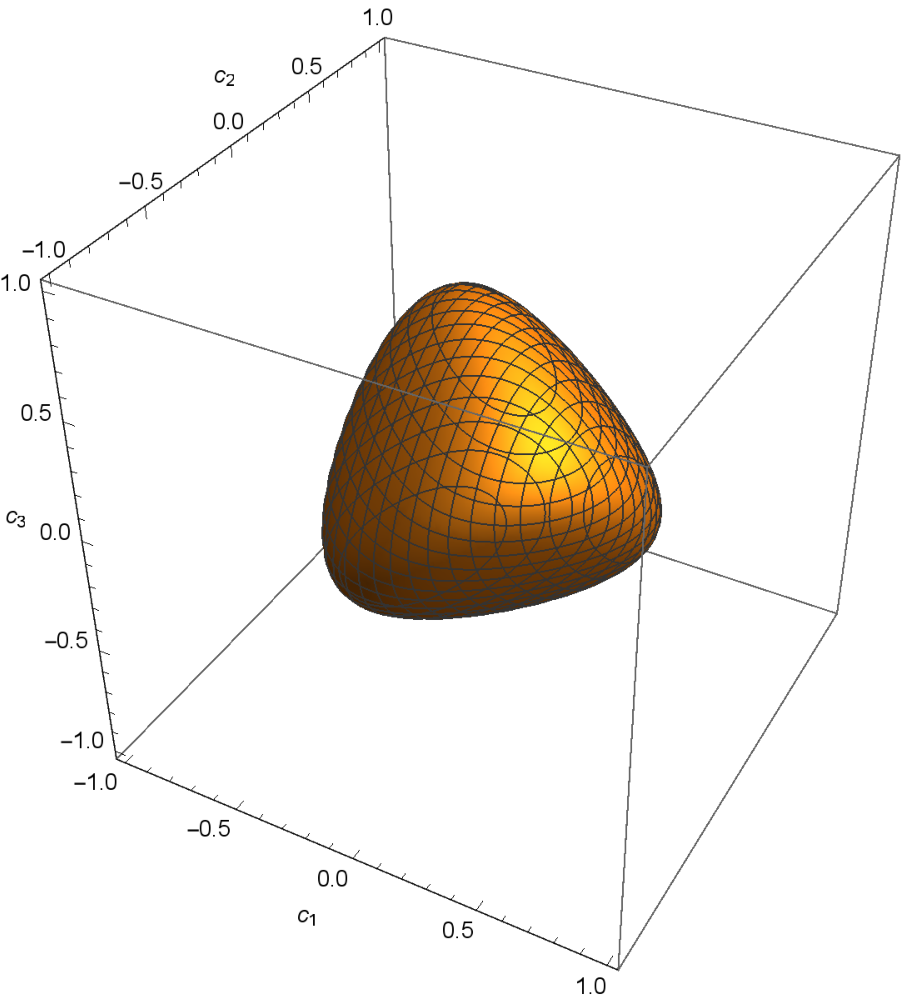}
\end{minipage}}
\subfigure[] {\begin{minipage}[b]{0.3\linewidth}
\includegraphics[width=1\textwidth]{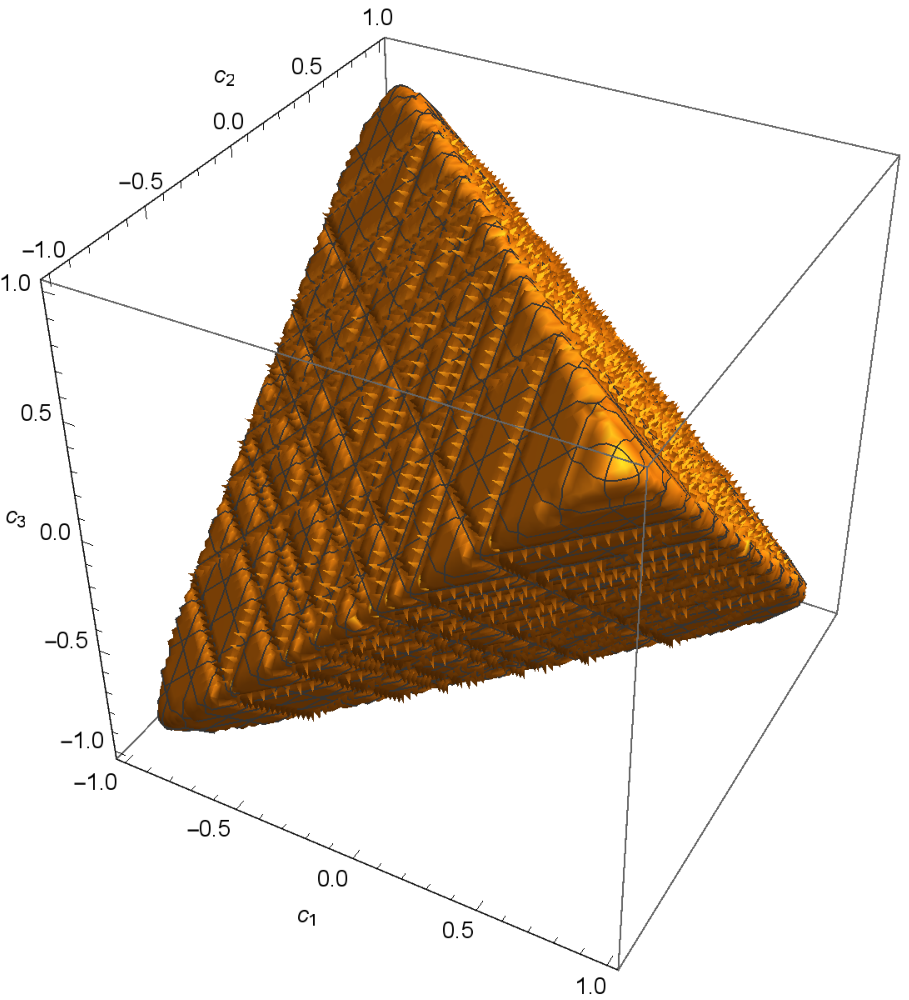}
\end{minipage}}
\caption{Surfaces of constant $C_I(\rho^{BD})_{a}$: (a)
$C_I(\rho^{BD})_{a}=0.05$; (b) $C_I(\rho^{BD})_{a}=0.2$; (c)
$C_I(\rho^{BD})_{a}=1$.}
\label{fig:MUB3}
\end{figure}

Now, we consider another special class of two qubit $X$ states. By
taking $\mathbf{r}=(0,0,r)$ and $\mathbf{s}=(0,0,s)$, state
(\ref{eq10}) becomes the following one \cite{YKW3}
\begin{equation}\label{eq11}
\rho^X_z=\frac{1}{4}\left(I\otimes I+r\sigma_3\otimes I+I\otimes s\sigma_3+\sum_{i=1}^3 c_i\sigma_i\otimes \sigma_i\right),
\end{equation}
which can be written as the following matrix in basis $a_1$
$$(\rho^X_z)_{a_1}=\frac{1}{4} \left(
\begin{array}{cccc}
 1+r+s+c_3 & 0 & 0 & c_1-c_2 \\
 0 & 1+r-s-c_3 & c_1+c_2 & 0 \\
 0 & c_1+c_2 & 1-r+s-c_3 & 0 \\
c_1-c_2 & 0 & 0 & 1-r-s+c_3 \\
\end{array}
\right).$$

Direct computation shows that
\begin{eqnarray}\label{eq12}
&&C_I(\rho^X_z)_{a_1}
\nonumber\\
&=&1-\frac{1}{16}\frac{1}{(c_1+c_2)^2+(r-s)^2}\left[\left(\sqrt{1-c_3+\sqrt{(c_1+c_2)^2+(r-s)^2}}(r-s+\sqrt{(c_1+c_2)^2+(r-s)^2})\right.\right.
\nonumber\\
&&\left.+\sqrt{1-c_3-\sqrt{(c_1+c_2)^2+(r-s)^2}}(-r+s+\sqrt{(c_1+c_2)^2+(r-s)^2})\right)^2
\nonumber\\
&&+\left(\sqrt{1-c_3-\sqrt{(c_1+c_2)^2+(r-s)^2}}(r-s+\sqrt{(c_1+c_2)^2+(r-s)^2})\right.
\nonumber\\
&&\left.\left.+\sqrt{1-c_3+\sqrt{(c_1+c_2)^2+(r-s)^2}}(-r+s+\sqrt{(c_1+c_2)^2+(r-s)^2})\right)^2\right]
\nonumber\\
&&-\frac{1}{16}\frac{1}{(c_1-c_2)^2+(r+s)^2}\left[\left(\sqrt{1+c_3+\sqrt{(c_1-c_2)^2+(r+s)^2}}(-r-s+\sqrt{(c_1-c_2)^2+(r+s)^2})\right.\right.
\nonumber\\
&&\left.+\sqrt{1+c_3+\sqrt{(c_1-c_2)^2+(r+s)^2}}(r+s+\sqrt{(c_1-c_2)^2+(r+s)^2})\right)^2
\nonumber\\
&&+\left(\sqrt{1+c_3-\sqrt{(c_1-c_2)^2+(r+s)^2}}(r+s-\sqrt{(c_1-c_2)^2+(r+s)^2})\right.
\nonumber\\
&&\left.\left.-\sqrt{1+c_3+\sqrt{(c_1-c_2)^2+(r+s)^2}}(r+s+\sqrt{(c_1-c_2)^2+(r+s)^2})\right)^2\right].
\end{eqnarray}

The surfaces of constant $C_I(\rho^X_z)_{a_1}$ are shown in Figure
4. It can be seen that for $r=s$, the volume given by the surfaces
expands for larger coherence, see Figures 4(a) and (c) or (b) and
(d).
\begin{figure}[ht]\centering
\subfigure[] {\begin{minipage}[b]{0.35\linewidth}
\includegraphics[width=1\textwidth]{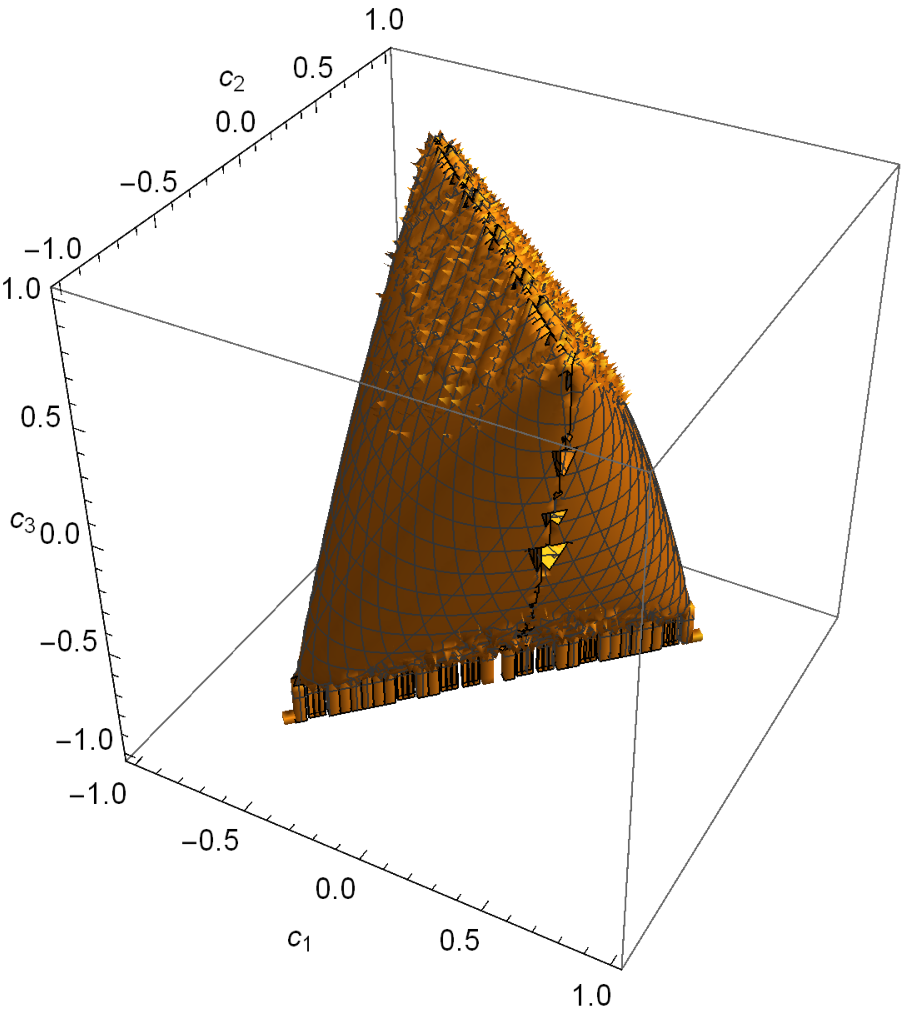}
\end{minipage}}
\subfigure[] {\begin{minipage}[b]{0.35\linewidth}
\includegraphics[width=1\textwidth]{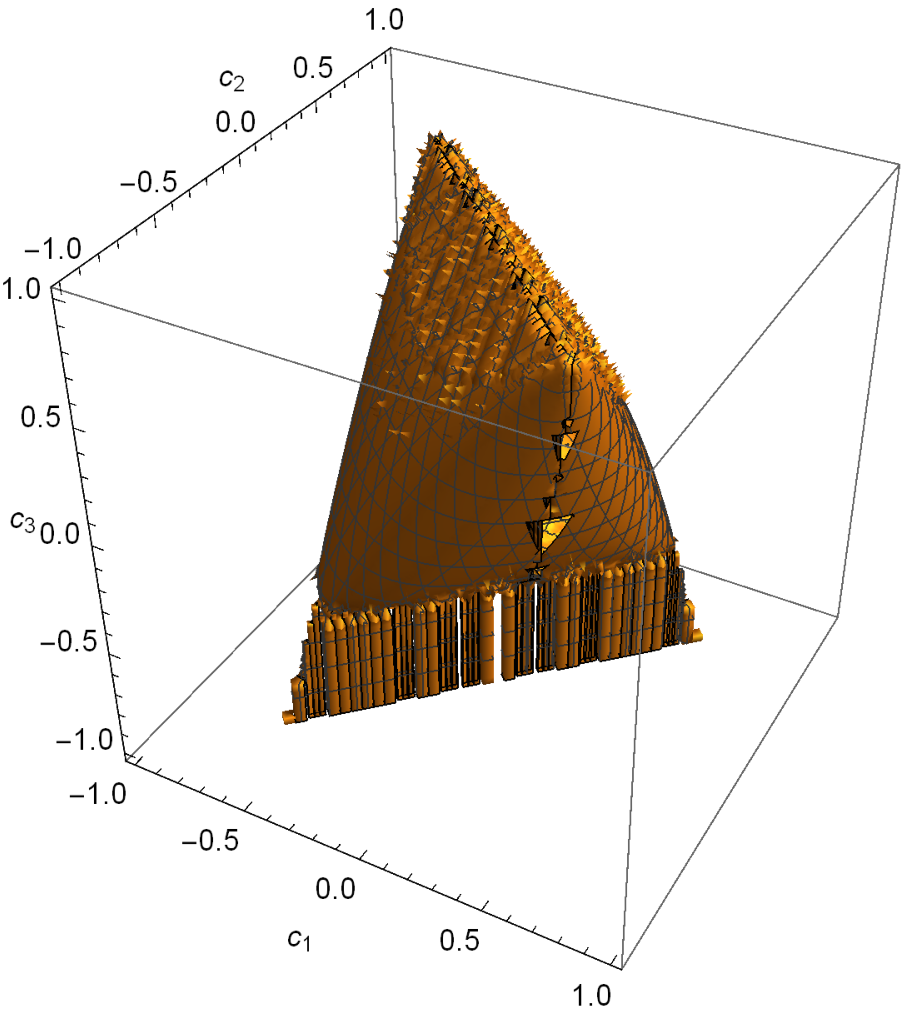}
\end{minipage}}
\subfigure[] {\begin{minipage}[b]{0.35\linewidth}
\includegraphics[width=1\textwidth]{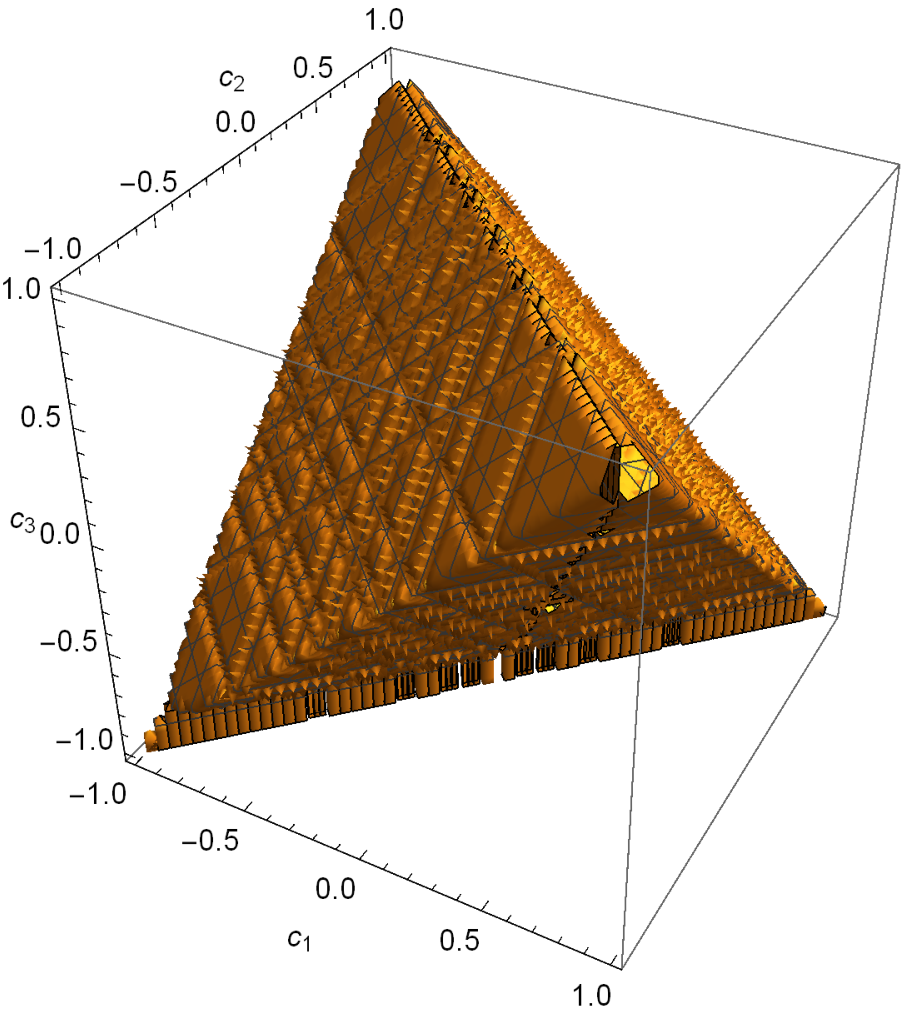}
\end{minipage}}
\subfigure[] {\begin{minipage}[b]{0.35\linewidth}
\includegraphics[width=1\textwidth]{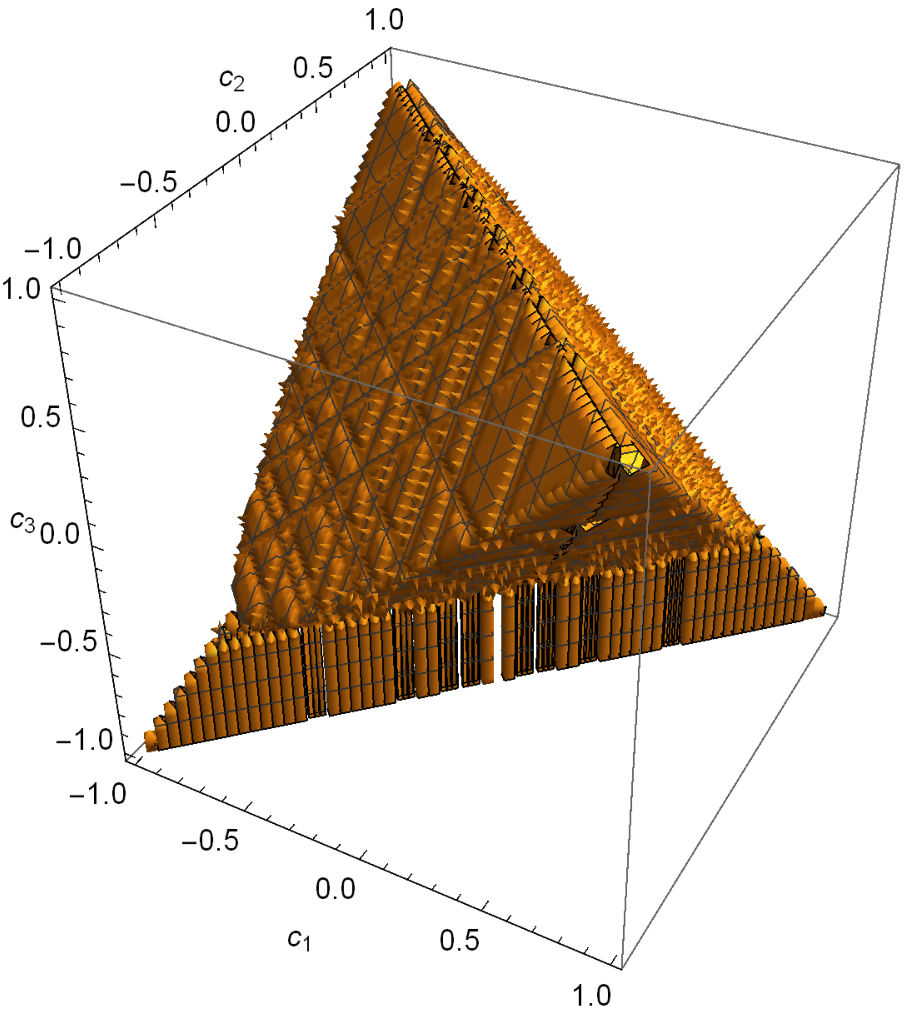}
\end{minipage}}
\caption{Surfaces of constant $C_I(\rho^X_z)_{a_1}$ with fixed $r$
and $s$: (a) $r=s=0.1,C_I(\rho^X_z)_{a_1}=0.1$; (b)
$r=s=0.3,C_I(\rho^X_z)_{a_1}=0.1$; (c)
$r=s=0.1,C_I(\rho^X_z)_{a_1}=0.5$; (d)
$r=s=0.3,C_I(\rho^X_z)_{a_1}=0.5$.} \label{fig:MUB4}
\end{figure}

Similarly, the matrix form of (\ref{eq11}) in basis $a_2$ and $a_3$
are
$$(\rho^X_z)_{a_2}=\frac{1}{4} \left(
\begin{array}{cccc}
 1+c_1 & s & r & c_3-c_2 \\
 s & 1-c_1 & c_2+c_3 & r \\
 r & c_2+c_3 & 1-c_1 & s \\
 c_3-c_2 & r & s & 1+c_1 \\
\end{array}
\right)$$
and
$$(\rho^X_z)_{a_3}=\frac{1}{4} \left(
\begin{array}{cccc}
 1+c_2 & s & r & c_3-c_1 \\
 s & 1-c_2 & c_1+c_3 & r \\
 r & c_1+c_3 & 1-c_2 & s \\
 c_3-c_1 & r & s & 1+c_2 \\
\end{array}
\right),$$ respectively, and $C_I(\rho^X_z)_{a_2}$ and
$C_I(\rho^X_z)_{a_3}$ can be similarly calculated.

Moreover, denoting the sum of the skew information-based coherence
of $\rho^X_z$ in bases $\{a_i\}_{i=1}^3$ by
\begin{equation}\label{eq14}
C_I(\rho^X_z)_{a}=C_I(\rho^X_z)_{a_1}+C_I(\rho^X_z)_{a_2}+C_I(\rho^{BD})_{a_3},
\end{equation}
we obtain that
\begin{eqnarray}
&&C_I(\rho^X_z)_{a}
\nonumber\\
&=&\frac{1}{4}\left(6-\sqrt{1-\sqrt{\left(c_1+c_2\right){}^2+(r-s)^2}-c_3}
\sqrt{1+\sqrt{\left(c_1+c_2\right){}^2+(r-s)^2}-c_3}\right.
\nonumber\\
&&\left.-\sqrt{1-\sqrt{\left(c_1-c_2\right){}^2+(r+s)^2}+c_3}
\sqrt{1+\sqrt{\left(c_1+c_2\right){}^2+(r-s)^2}-c_3}\right.
\nonumber\\
&&\left.\sqrt{1-\sqrt{\left(c_1-c_2\right){}^2+(r+s)^2}+c_3}
\sqrt{1+\sqrt{\left(c_1+c_2\right){}^2+(r-s)^2}-c_3}\right.
\nonumber\\
&&\left.-\sqrt{1-\sqrt{\left(c_1+c_2\right){}^2+(r-s)^2}-c_3}
\sqrt{1-\sqrt{\left(c_1-c_2\right){}^2+(r+s)^2}+c_3}\right.
\nonumber\\
&&\left.-\sqrt{1-\sqrt{\left(c_1+c_2\right){}^2+(r-s)^2}-c_3}
\sqrt{1+\sqrt{\left(c_1-c_2\right){}^2+(r+s)^2}+c_3}\right.
\nonumber\\
&&\left.-\sqrt{1-\sqrt{\left(c_1-c_2\right){}^2+(r+s)^2}+c_3}
\sqrt{1+\sqrt{\left(c_1-c_2\right){}^2+(r+s)^2}+c_3}\right),
\end{eqnarray}
and the surfaces of constant $C_I(\rho^X_z)_{a}$ are shown in Figure
5. It can be seen that similar properties hold compared with the
surfaces of constant $C_I(\rho^X_z)_{a_1}$ in Figure 4.

\begin{figure}[ht]\centering
\subfigure[] {\begin{minipage}[b]{0.35\linewidth}
\includegraphics[width=1\textwidth]{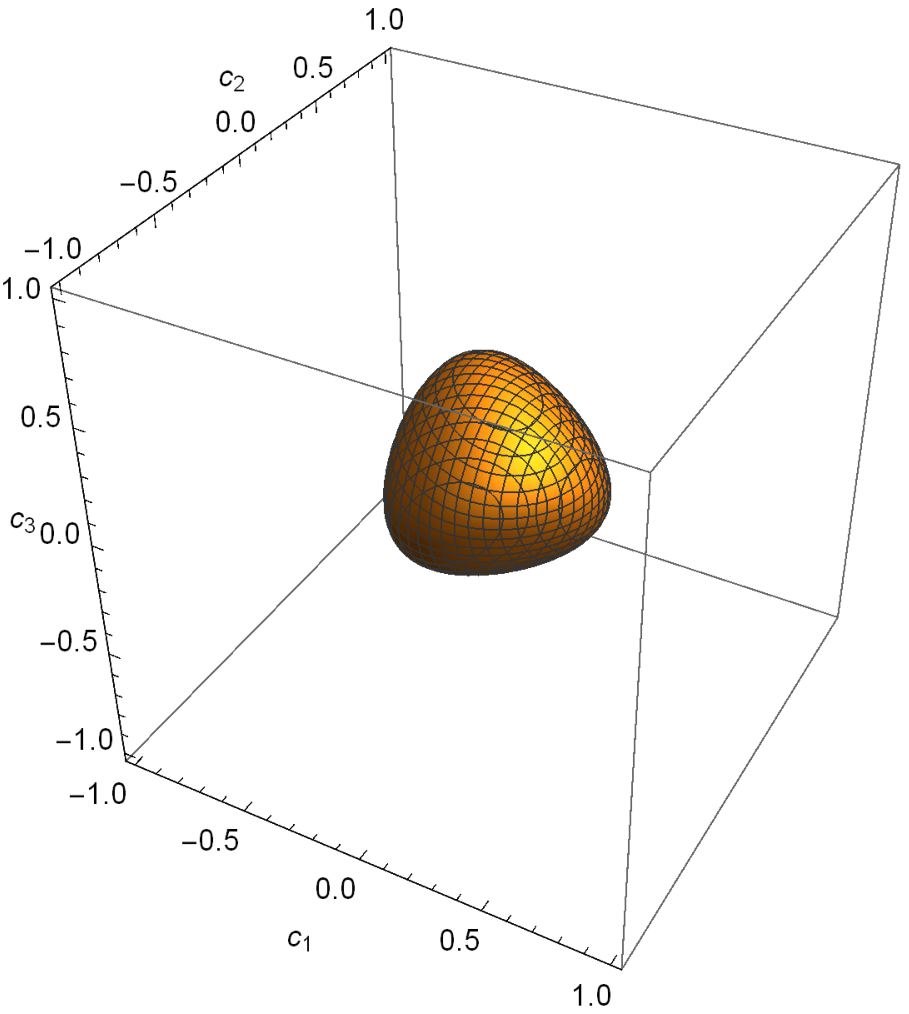}
\end{minipage}}
\subfigure[] {\begin{minipage}[b]{0.35\linewidth}
\includegraphics[width=1\textwidth]{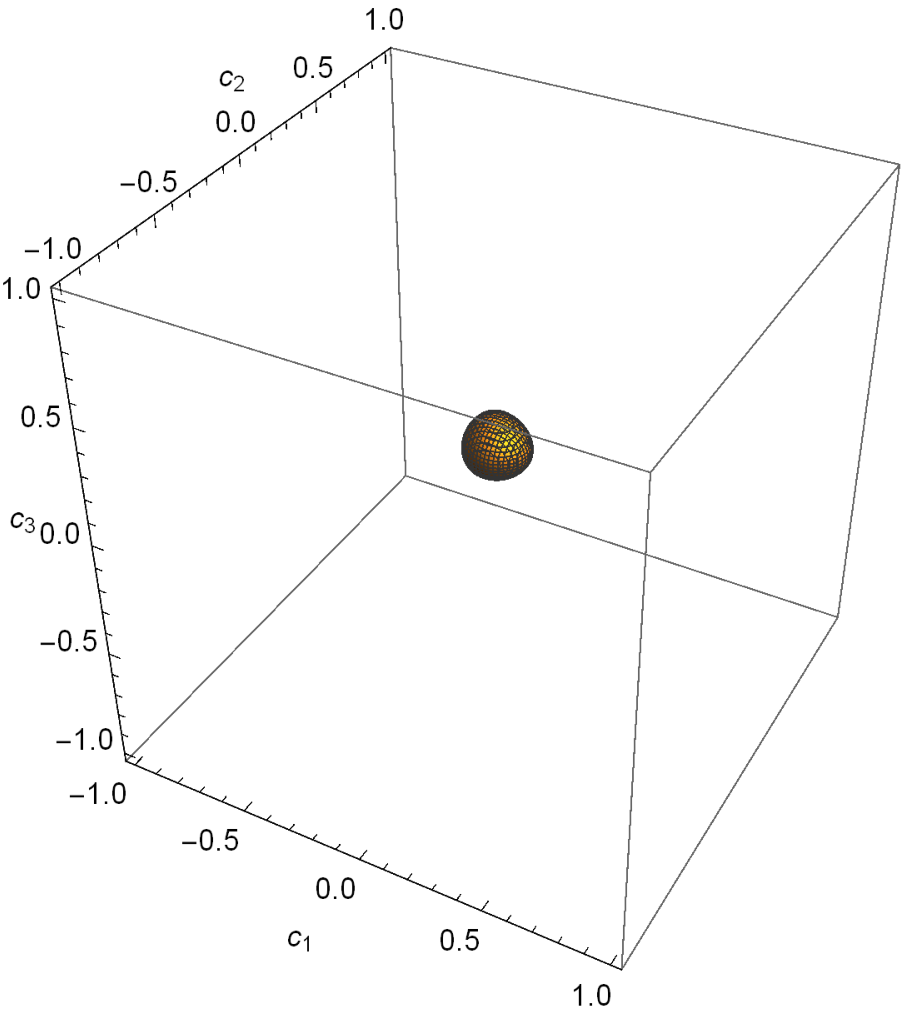}
\end{minipage}}
\subfigure[] {\begin{minipage}[b]{0.35\linewidth}
\includegraphics[width=1\textwidth]{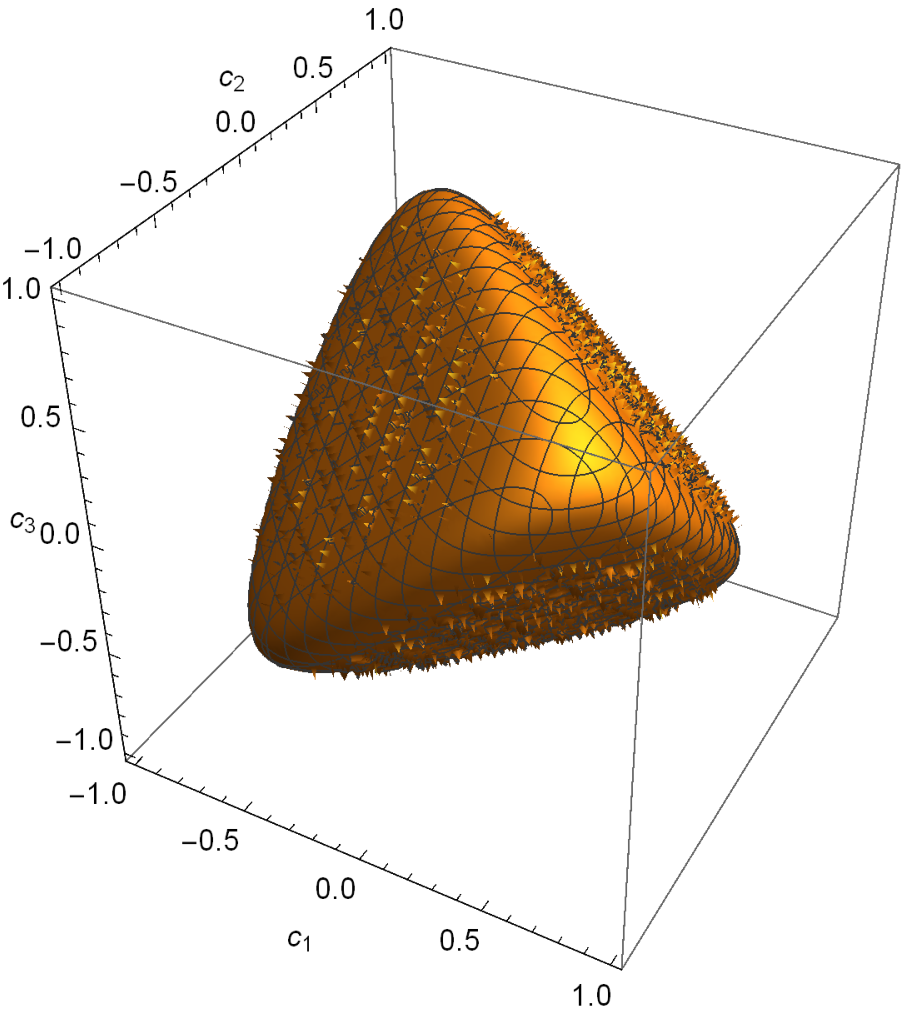}
\end{minipage}}
\subfigure[] {\begin{minipage}[b]{0.35\linewidth}
\includegraphics[width=1\textwidth]{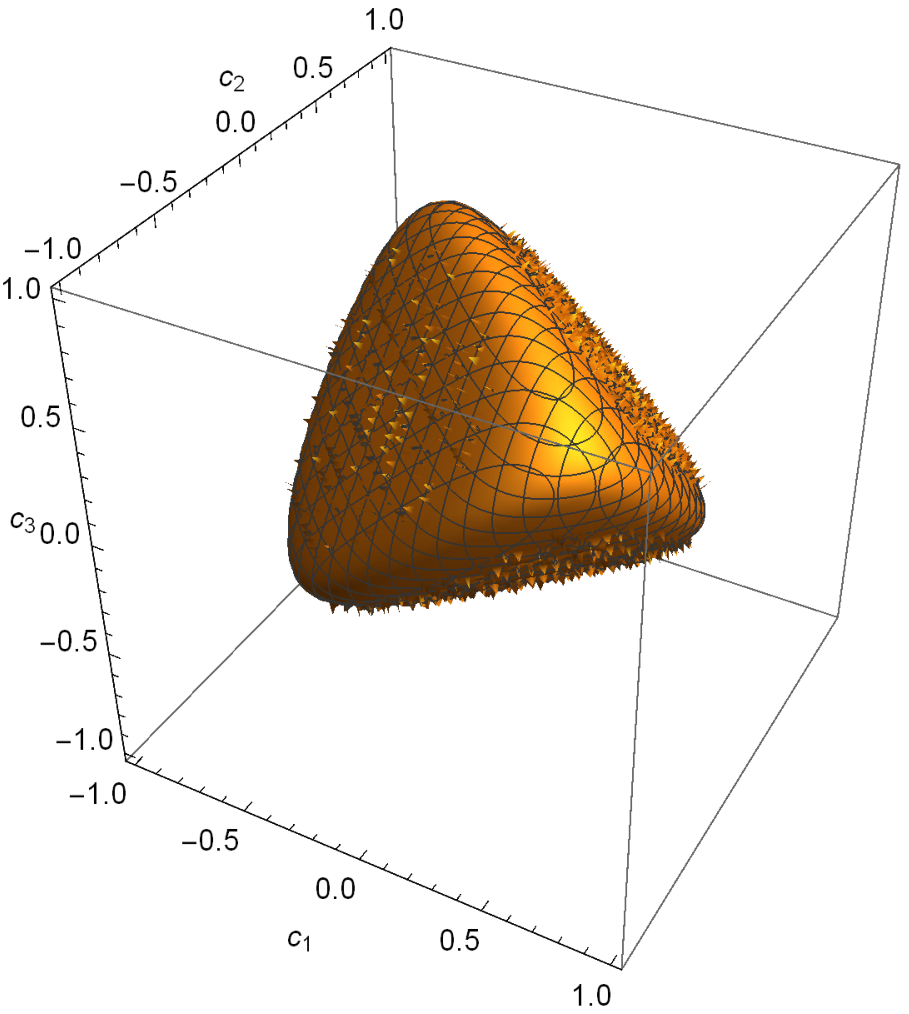}
\end{minipage}}
\caption{Surfaces of constant $C_I(\rho^X_z)_{a}$ with fixed $r$ and
$s$: (a) $r=s=0.1,C_I(\rho^X_z)_{a}=0.1$; (b)
$r=s=0.3,C_I(\rho^X_z)_{a}=0.1$; (c)
$r=s=0.1,C_I(\rho^X_z)_{a}=0.5$; (d)
$r=s=0.3,C_I(\rho^X_z)_{a}=0.5$.} \label{fig:MUB5}
\end{figure}

\noindent {\bf 3. Skew information-based coherence under quantum
channels}

We now consider the evolution of the skew information-based quantum
coherence under different quantum channels. Consider the
following type of quantum channel $\Phi$:
\begin{equation}\label{eq15}
\Phi(\rho)=\sum_{i,j}(E_i\otimes E_j)\rho(E_i\otimes E_j)^\dag,
\end{equation}
where $\{E_k\}$ is the set of Kraus operators on a single qubit,
satisfying $\sum_k E_k^\dag E_k=I$. The Kraus operators for four
kinds of quantum channels are listed in Table 1 \cite{MPSS}.
{\begin{table} \caption{  {Kraus operators for the quantum channels:
bit flip (BF), phase flip (PF), bit-phase flip (BPF), and
generalized amplitude damping (GAD), where $p$ and $\gamma$ are
decoherence probabilities, $0<p<1$, $0<\gamma<1$.} }
$$\begin{array}{ccccc}\hline\hline
\text{Channel}\hspace{1.3in}   &  \text{Kraus operators}\\ \hline
 \mathrm{BF}\hspace{1.3in}   &   E_0=\sqrt{1-p/2}I,\,\,\,\,\,\,  E_1=\sqrt{p/2}\sigma_1  \\
 \mathrm{PF}\hspace{1.3in}   &   E_0=\sqrt{1-p/2}I,\,\,\,\,\,\,  E_1=\sqrt{p/2}\sigma_3  \\
 \mathrm{BPF}\hspace{1.3in}  &   E_0=\sqrt{1-p/2}I,\,\,\,\,\,\,  E_1=\sqrt{p/2}\sigma_2  \\
 \mathrm{GAD}\hspace{1.3in}  &   E_0=\sqrt{p}\left(\begin{array}{cc}
         1&0\\
         0&\sqrt{1-\gamma}\\
         \end{array}
         \right), \,\,\,\,\,\, E_2=\sqrt{1-p}\left(\begin{array}{cc}
         \sqrt{1-\gamma}&0\\
         0&1\\
         \end{array}
         \right)  \\
  &  E_1=\sqrt{p}\left(\begin{array}{cc}
         0&\sqrt{\gamma}\\
         0&0\\
         \end{array}
         \right), \,\,\,\,\,\, E_3=\sqrt{1-p}\left(\begin{array}{cc}
         0&0\\
         \sqrt{\gamma}&0\\
         \end{array}
         \right)\\ \hline\hline
\end{array}$$
\end{table}}

Noting that a Bell-diagonal state under BF, PF and BPF in Table 1
remains the same form, which is also the case under GAD for $p=1/2$
and any $\gamma$, we have
\begin{equation}\label{eq16}
\Phi(\rho^{BD})=\frac{1}{4}(I\otimes I+\sum_{i=1}^3
c_i'\sigma_i\otimes \sigma_i),
\end{equation}
where $\rho^{BD}$ is a two-qubit Bell-diagonal state, and the
parameters $c_i'$ $(i=1,2,3)$ are listed in Table 2 \cite{MPSS}.
{\begin{table} \caption{  {Correlation coefficients with respect to
the following channels: bit flip (BF), phase flip (PF), bit-phase
flip (BPF), and generalized amplitude damping (GAD). For GAD, we
have fixed $p=1/2$ and replaced $\gamma$ by $p$.} }
$$\begin{array}{ccccc}\hline\hline
\text{Channel}\hspace{0.9in} & \text{$c_1'$} \hspace{0.9in} & \text{$c_2'$} \hspace{0.9in} & \text{$c_3'$}\\
\hline
 \mathrm{BF}\hspace{0.9in}  &  c_1        \hspace{0.9in}  & c_2(1-p)^2  \hspace{0.9in} &  c_3(1-p)^2\\
 \mathrm{PF}\hspace{0.9in}  &  c_1(1-p)^2 \hspace{0.9in}  & c_2(1-p)^2  \hspace{0.9in} &  c_3\\
 \mathrm{BPF}\hspace{0.9in} &  c_1(1-p)^2 \hspace{0.9in}  & c_2         \hspace{0.9in} &  c_3(1-p)^2\\
 \mathrm{GAD}\hspace{0.9in} &  c_1(1-p)   \hspace{0.9in}  & c_2(1-p)    \hspace{0.9in} &  c_3(1-p)^2\\ \hline\hline
\end{array}$$
\end{table}}

By replacing $c_i$ by $c_i'$ in Eq. (\ref{eq11}), we plot the
surfaces of constant $C_I(\Phi(\rho^{BD}))_{a_1}$ for the four types
of channels by utilizing Table 2, see Figures 6, 7, 8 and 9. For
simplicity, we use $C_{BF}$, $C_{PF}$, $C_{BPF}$ and $C_{GAD}$ to
represent $C_I(\Phi(\rho^{BD}))_{a_1}$, where $\Phi$ is BF, PF, BPF
and GAD, respectively. The surfaces show interesting shapes for
parameter $p$ and the coherence $C$. When both $p$ and $C$ are
small, the surface is very similar for four channels, see Figures
6-9(a). When $p$ is small and $C$ is large, the surface is two
separate pieces of a tetrahedron with a gap in different directions
for BF, PF and BPF channels, see Figures 6-8(b), and four pieces of
a tetrahedron, see Figure 9(b). When $p$ is large and $C$ is small,
the surface is two opposite surfaces for BF, PF and BPF channels,
see Figures 6-8(c), and is four pieces of surfaces, of which two
pairs are opposite for GAD channels, see Figure 9(c).
\begin{figure}[ht]\centering
\subfigure[] {\begin{minipage}[b]{0.3\linewidth}
\includegraphics[width=1\textwidth]{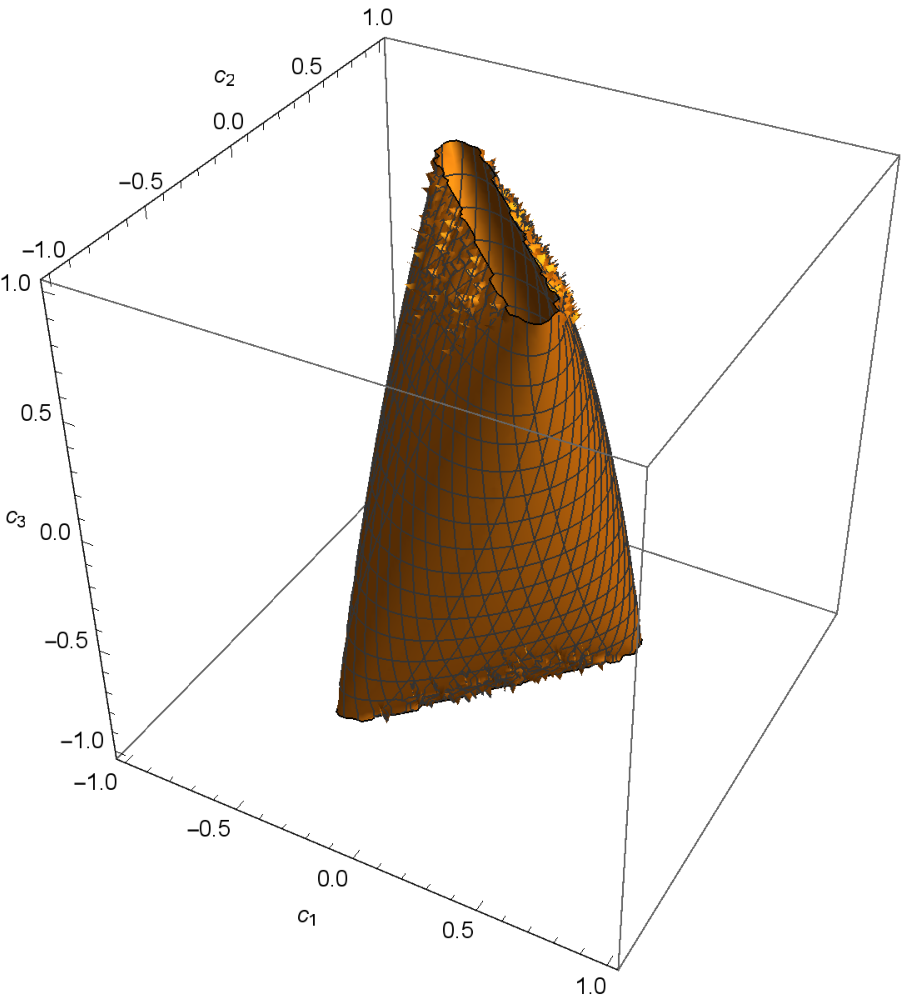}
\end{minipage}}
\subfigure[] {\begin{minipage}[b]{0.3\linewidth}
\includegraphics[width=1\textwidth]{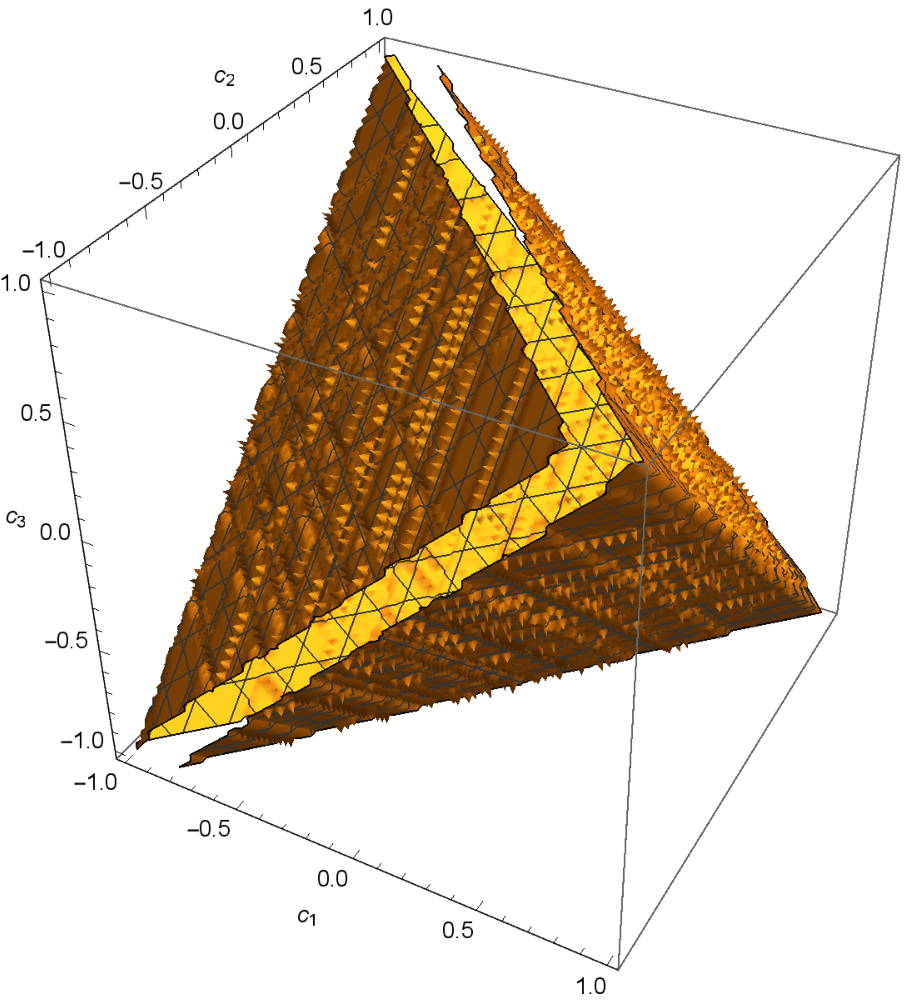}
\end{minipage}}
\subfigure[] {\begin{minipage}[b]{0.3\linewidth}
\includegraphics[width=1\textwidth]{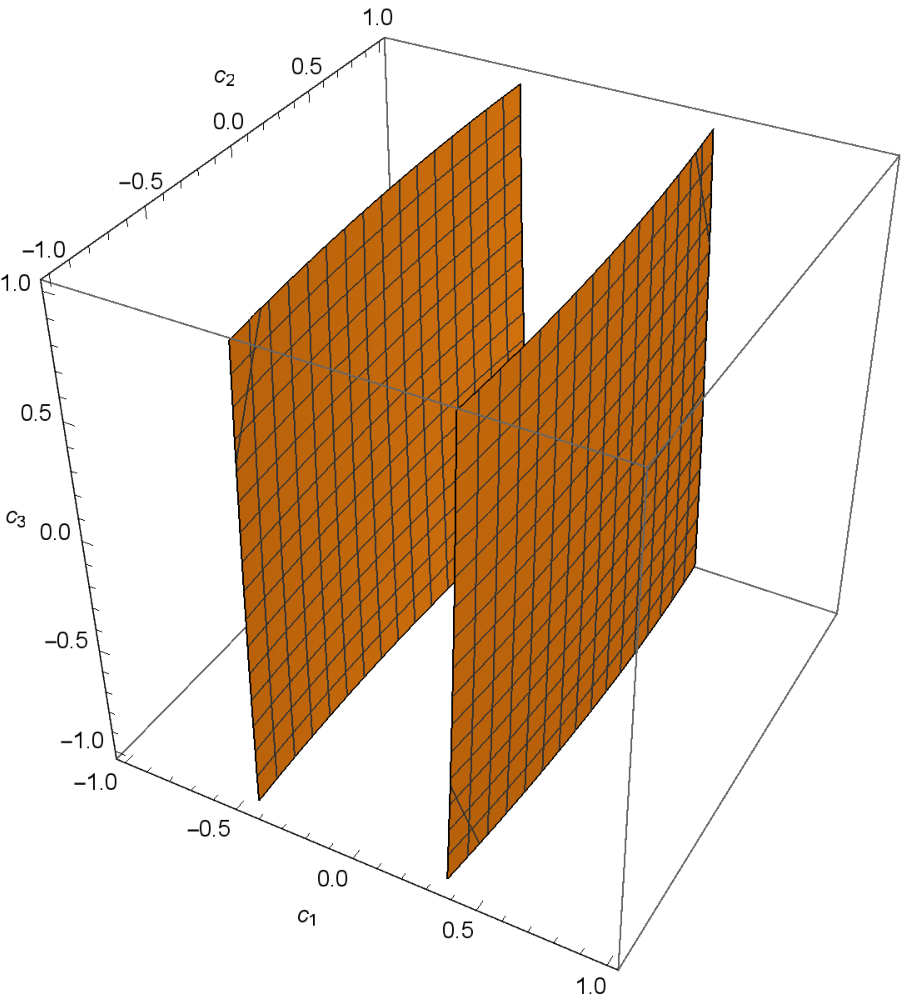}
\end{minipage}}
\caption{Surfaces of constant $C_{BF}$ for bit flip channels with
fixed $p$: (a) $p=0.05,C_{BF}=0.05$; (b) $p=0.05,C_{BF}=0.4$; (c)
$p=0.6,C_{BF}=0.05$.} \label{fig:MUB6}
\end{figure}

\begin{figure}[ht]\centering
\subfigure[] {\begin{minipage}[b]{0.3\linewidth}
\includegraphics[width=1\textwidth]{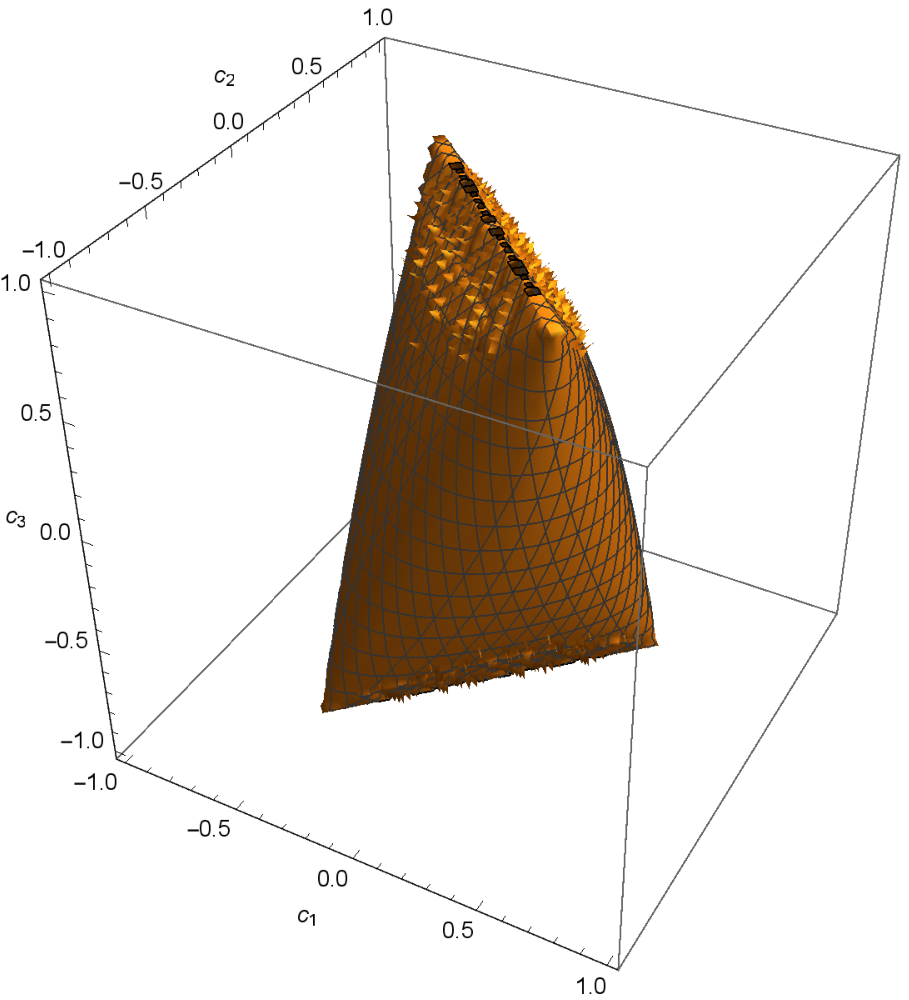}
\end{minipage}}
\subfigure[] {\begin{minipage}[b]{0.3\linewidth}
\includegraphics[width=1\textwidth]{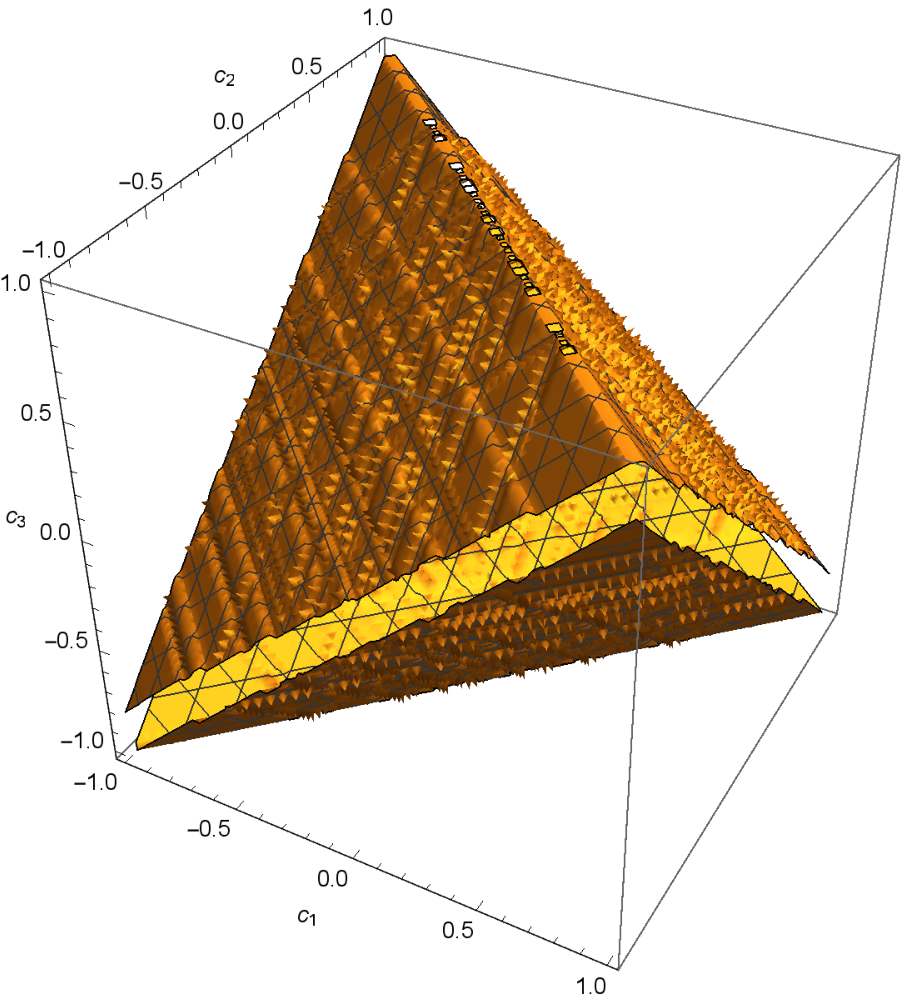}
\end{minipage}}
\subfigure[] {\begin{minipage}[b]{0.3\linewidth}
\includegraphics[width=1\textwidth]{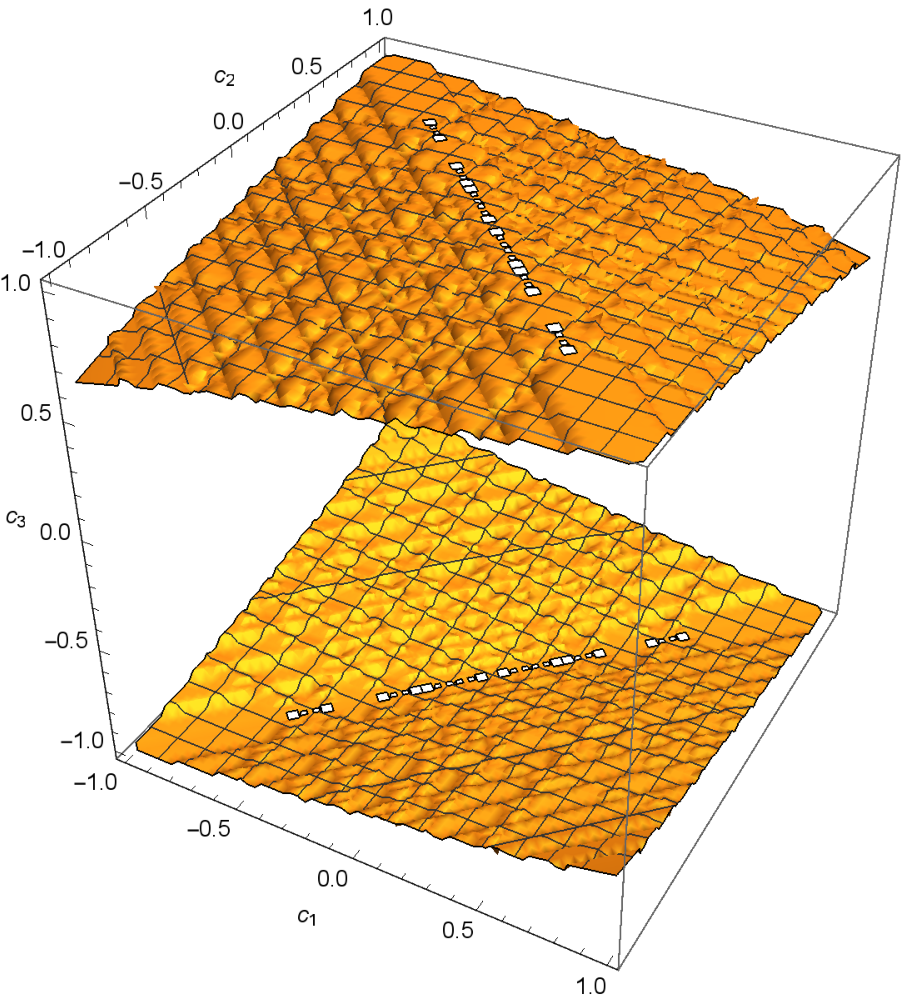}
\end{minipage}}
\caption{Surfaces of constant $C_{PF}$ for phase flip channels with
fixed $p$: (a) $p=0.05,C_{PF}=0.05$; (b) $p=0.05,C_{PF}=0.4$; (c)
$p=0.6,C_{PF}=0.05$.} \label{fig:MUB7}
\end{figure}

\begin{figure}[ht]\centering
\subfigure[] {\begin{minipage}[b]{0.3\linewidth}
\includegraphics[width=1\textwidth]{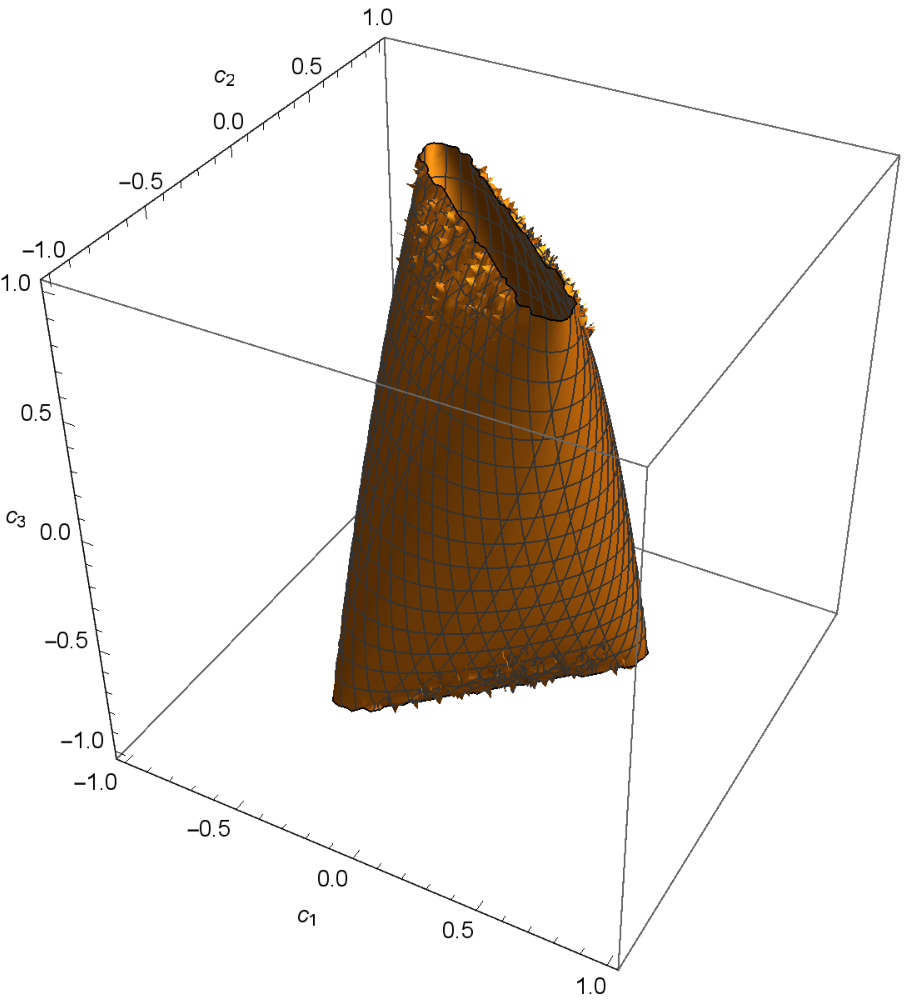}
\end{minipage}}
\subfigure[] {\begin{minipage}[b]{0.3\linewidth}
\includegraphics[width=1\textwidth]{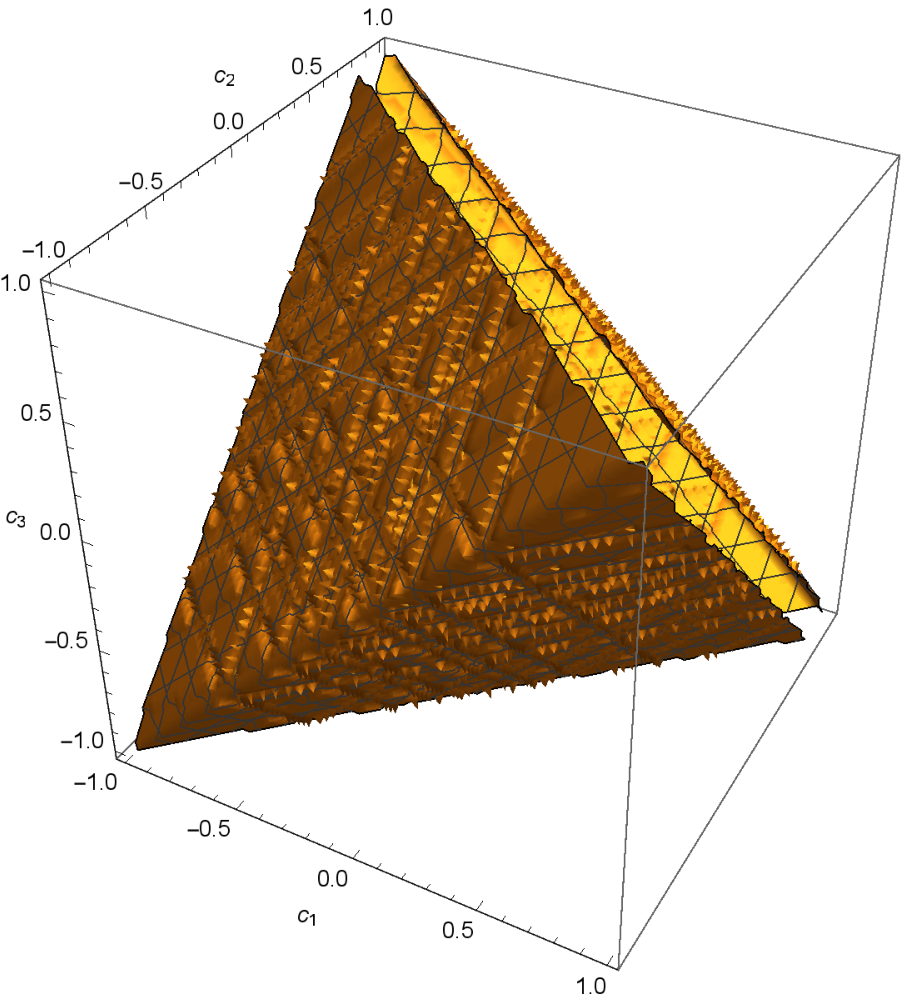}
\end{minipage}}
\subfigure[] {\begin{minipage}[b]{0.3\linewidth}
\includegraphics[width=1\textwidth]{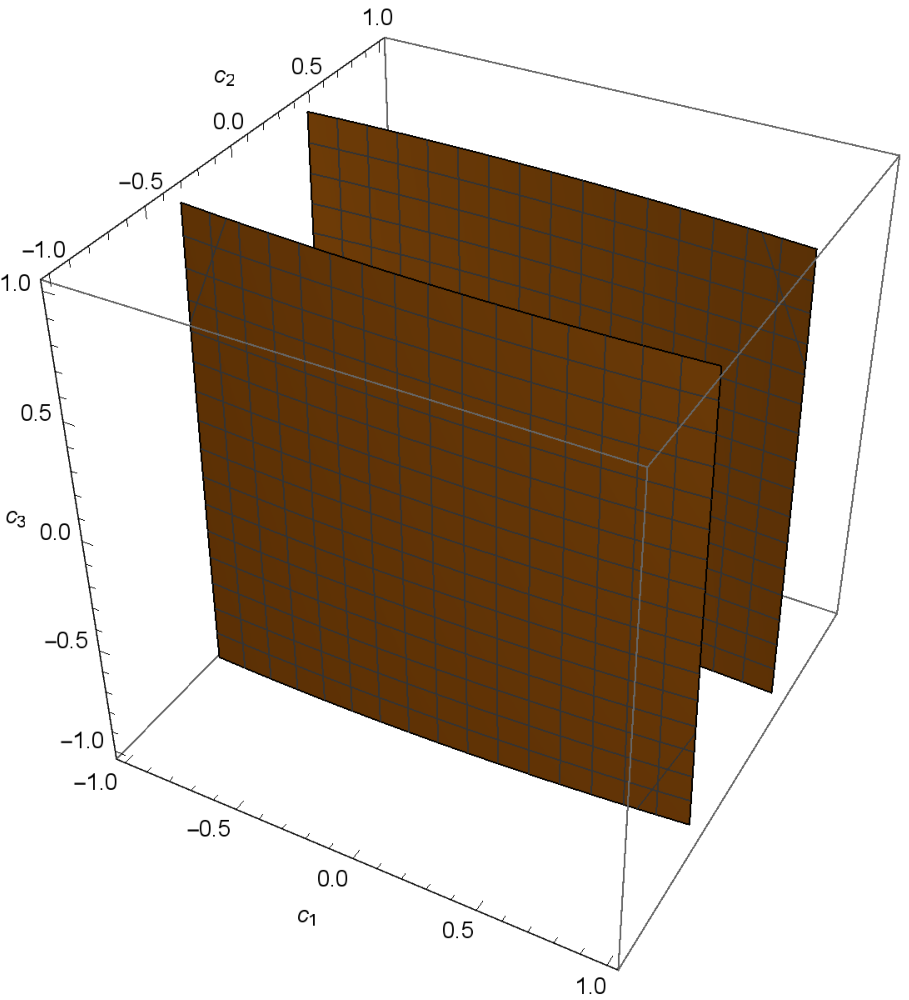}
\end{minipage}}
\caption{Surfaces of constant $C_{BPF}$ for bit-phase flip channels
with fixed $p$: (a) $p=0.05,C_{BPF}=0.05$; (b) $p=0.05,C_{BPF}=0.4$;
(c) $p=0.6,C_{BPF}=0.05$.} \label{fig:MUB8}
\end{figure}

\begin{figure}[ht]\centering
\subfigure[] {\begin{minipage}[b]{0.3\linewidth}
\includegraphics[width=1\textwidth]{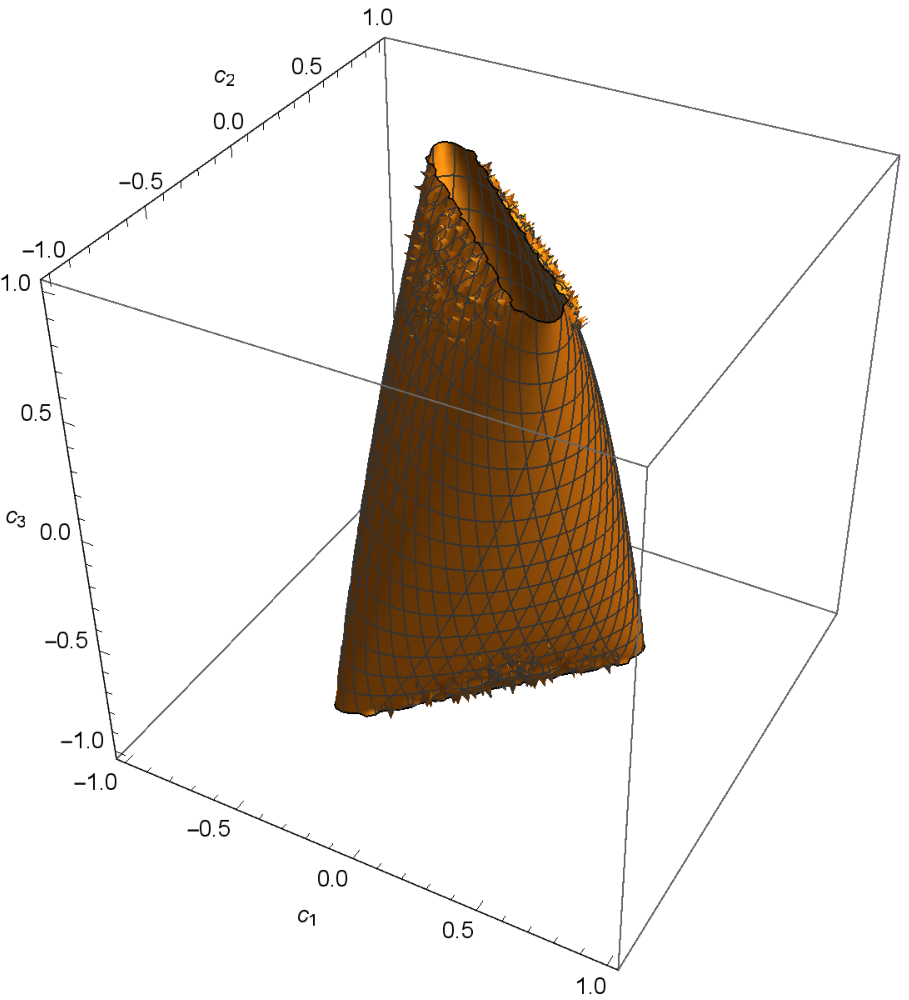}
\end{minipage}}
\subfigure[] {\begin{minipage}[b]{0.3\linewidth}
\includegraphics[width=1\textwidth]{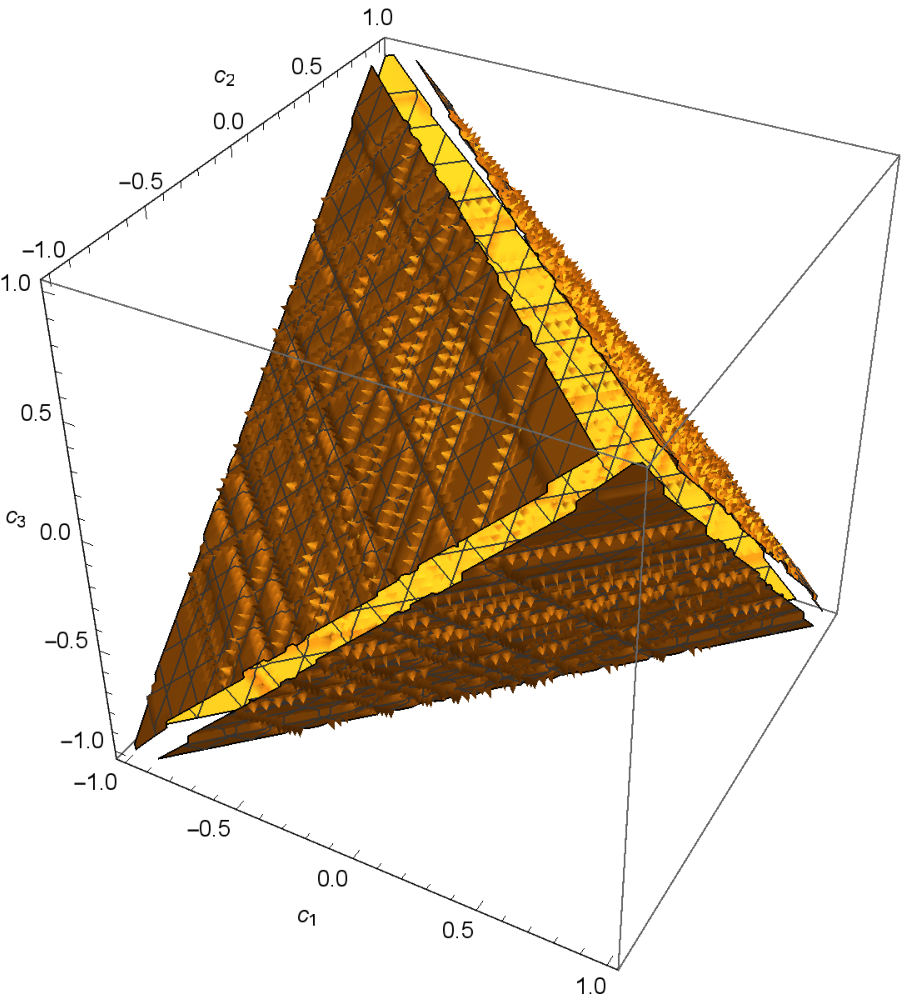}
\end{minipage}}
\subfigure[] {\begin{minipage}[b]{0.3\linewidth}
\includegraphics[width=1\textwidth]{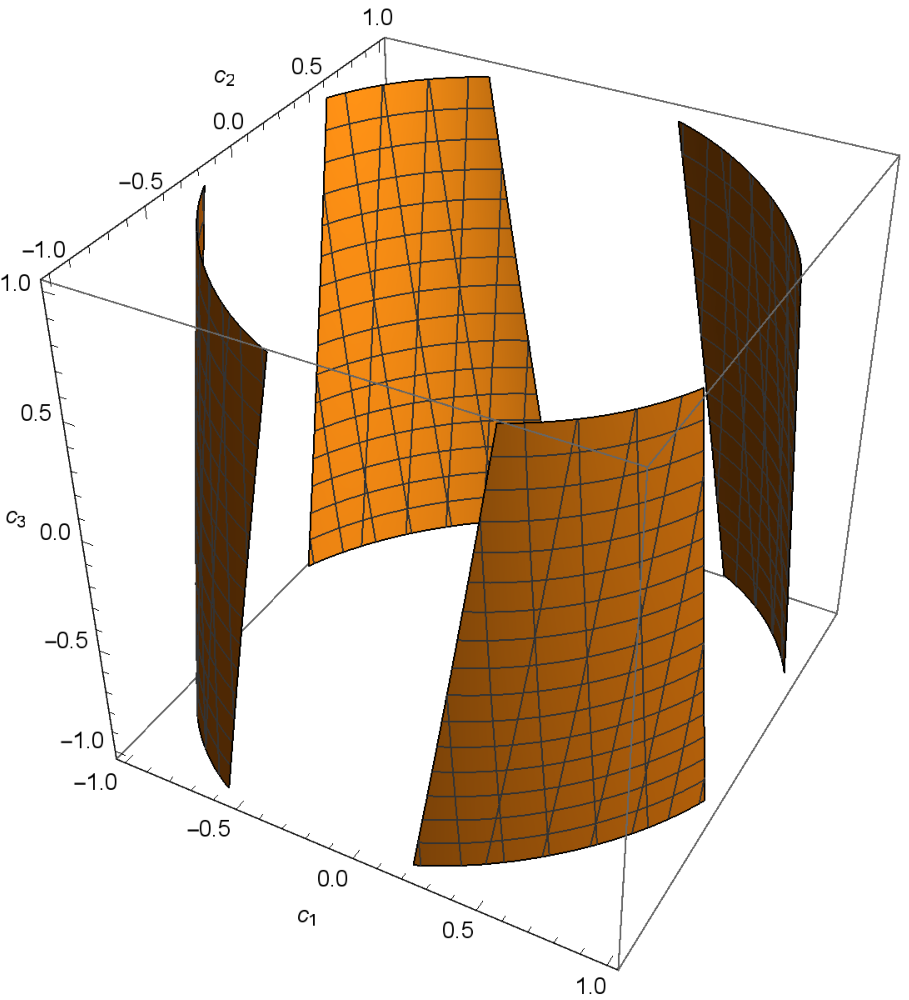}
\end{minipage}}
\caption{Surfaces of constant $C_{GAD}$ for generalized amplitude
damping channels with fixed $p$: (a) $p=0.05,C_{GAD}=0.05$; (b)
$p=0.05,C_{GAD}=0.4$; (c) $p=0.6,C_{GAD}=0.05$.} \label{fig:MUB9}
\end{figure}

Setting $c_1=-0.2, c_2=0.6, c_3=0.6$ and $c_1=-0.6,c_2=0.2,c_3=0.2$,
respectively, the dynamics of $C_I(\rho^{BD})$ under BF, PF, BPF and
GAD channels are shown in Figure 10. The $C$-axis denotes $C_{BF},
C_{PF}, C_{BPF}$ and $C_{GAD}$. Similar to the case of relative
entropy of coherence in \cite{YKW4}, we find that all the curves are
decreasing functions of $p$, and $C_{PF}$ and $C_{GAD}$ approaches
to zero as $p$ approaches $1$. Moreover, $C_{BF}$ decreases
dramatically as $p$ increases, see Figure 10(a), while it decreases
very slowly in case of Figure 10(b).
\begin{figure}[ht]\centering
\subfigure[] {\begin{minipage}[b]{0.47\linewidth}
\includegraphics[width=1\textwidth]{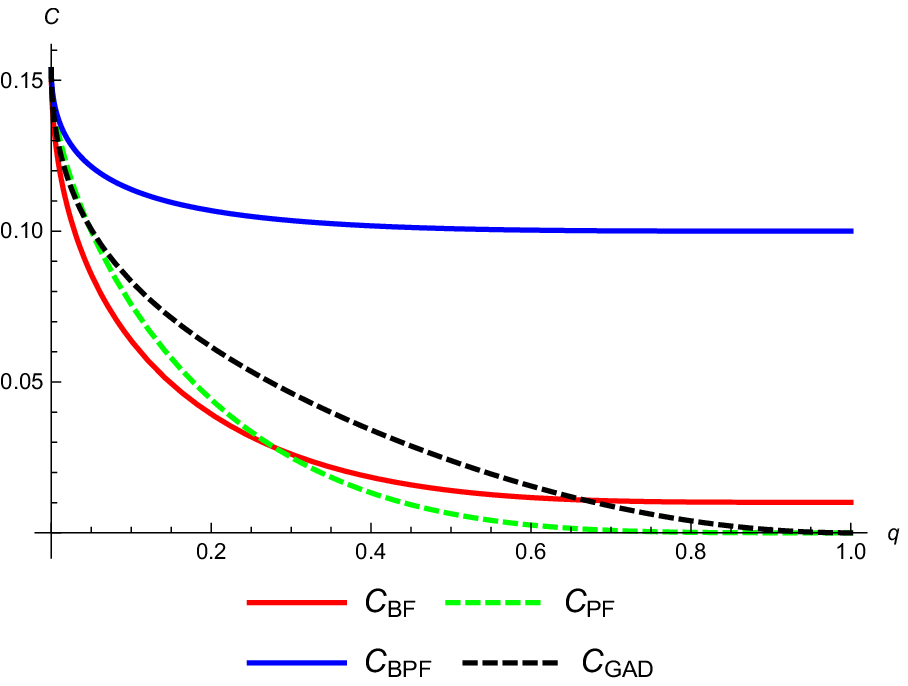}
\end{minipage}}
\subfigure[] {\begin{minipage}[b]{0.47\linewidth}
\includegraphics[width=1\textwidth]{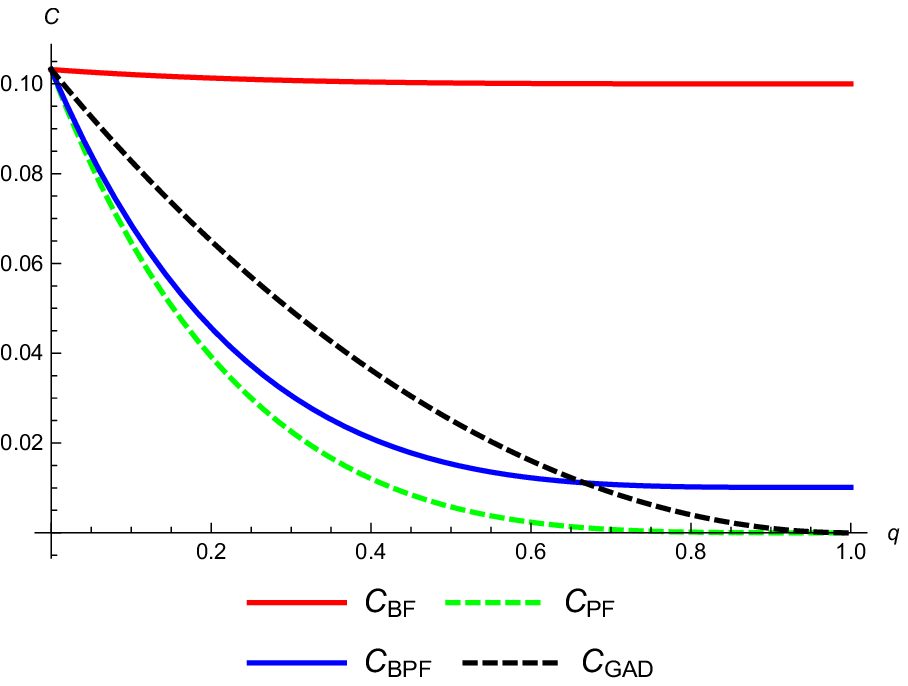}
\end{minipage}}
\caption{$C_{BF},C_{PF},C_{BPF}$ and $C_{GAD}$ as a function of $p$:
(a) $c_1=-0.2, c_2=0.6, c_3=0.6$; (b) $c_1=-0.6,c_2=0.2,c_3=0.2$.}
\label{fig:MUB10}
\end{figure}

\vskip0.1in

\noindent {\bf 4. Conclusions and discussions}

\vskip0.1in

We have studied the geometry of skew information-based coherence of
quantum states in mutually unbiased bases by calculating the skew
information-based coherence for two qubit states. We calculated the
analytical expression for Bell-diagonal states and a special class
of $X$ states in a set of autotensor of mutually unbiased bases. As
direct consequences, explicit formulas for coherences of Werner
states and isotropic states have also been given, which shows that
the former is an decreasing function of the state parameter, while
the latter is not. Based on these, the geometry of skew
information-based coherence for these two qubit states in both
computational basis and in MUBs have been depicted.

Moreover, we have displayed the level surfaces of skew
information-based coherence for Bell-diagonal states under four
typical local nondissipative quantum channels. It has been shown
that similar trend occurs when the relative entropy of coherence is
used, but the shape of the graphics turned out to be very different.
Furthermore, by choosing two different sets of fixed values for
$c_1$, $c_2$ and $c_3$, we have depicted the skew information-based
coherence under the four channels as a function of the parameter
$p$. It shows the similar features as the relative entropy of
coherence.

\vskip0.1in

\noindent

\subsubsection*{Acknowledgements}
This work was supported by National Natural Science Foundation of
China (11701259, 11461045, 11675113), the China Scholarship Council
(201806825038), the Key Project of Beijing Municipal Commission of
Education (KZ201810028042), Beijing Natural Science Foundation
(Z190005), and Academy for Multidisciplinary Studies, Capital Normal University.
This work was completed while Zhaoqi Wu was visiting
Max-Planck-Institute for Mathematics in the Sciences in Germany.


\end{document}